# Light-Driven Ultrafast Control of Time-Reversal Symmetry in $SrTiO_3$

In Hyeok Choi[1,†], Sergei Urazhdin[2], Shivasheesh Varshney[3], Seung Gyo Jeong[3,4], Bharat Jalan[3] and Keith A. Nelson[1,*]

[1]*Department of Chemistry, Massachusetts Institute of Technology, Cambridge, Massachusetts 02139, United States*

[2]*Department of Physics, Emory University, Atlanta, Georgia 30322, United States*

[3]*Department of Chemical Engineering and Materials Science, University of Minnesota−Twin Cities, Minneapolis, Minnesota 55455, United States*

[4]*Department of Physics, Hankuk University of Foreign Studies, Yongin 17035, Republic of Korea*

[†] First author

[*]Corresponding authors: kanelson@mit.edu

**Abstract**

Light-driven control of material symmetry enables the engineering of phenomena forbidden in equilibrium. Although electromagnetic fields have been used to break time-reversal symmetry, its dynamic control on ultrafast timescales in nonmagnetic insulating oxides remains virtually unexplored. Here, we demonstrate simultaneous magnetic symmetry lowering and time-reversal-symmetry breaking in cubic $SrTiO_3$ driven by an off-resonant, elliptically polarized THz pulse. Second harmonic generation (SHG) polarimetry reveals emergent SHG circular dichroism inconsistent with third-order nonlinear processes under cubic $m3m$ symmetry. The SHG polar patterns exhibit mirror-symmetry breaking, indicating a transient reduction to tetragonal $4/mm'm'$ magnetic symmetry. Furthermore, the SHG response scales linearly with the angular momentum of THz pulse, providing direct evidence for THz-field-induced time-reversal-symmetry breaking. These results establish a direct pathway for ultrafast symmetry control in non-magnetic oxides, enabling dynamic manipulation of emergent phases on ultrafast timescales.

Symmetry plays a crucial role in the physical properties of condensed-matter systems. According to Neumann's principle, macroscopic physical quantities must conform to the symmetry of materials, which is inherently determined by the periodic arrangement of ions, electron orbitals and spins. This fundamental concept provides the foundation for understanding nonlinear optical processes[1,2], as well as transport of charge, angular momentum and heat[3-9]. The spatial and time-reversal symmetries of materials also govern their electronic, structural, and magnetic phases, including superconductivity[10], ferroelectricity[9], ferromagnetism[11], and altermagnetism[12,13].

Control of symmetry through strain[14-19], doping[20-22], and optical excitation[23-27] has emerged as a transformative strategy for tailoring material properties beyond the intrinsic limits. In particular, coupling of terahertz (THz) light to the lattice enables ultrafast manipulation of crystal symmetry by directly exciting low-energy quasiparticles, resulting in structural and electronic phase transitions. Recent studies show that strong THz fields can break inversion symmetry in the low-temperature quantum paraelectric state of $SrTiO_3$ and $KTaO_3$[23,26], resulting in a macroscopic polarization. Furthermore, resonant excitation of mid-infrared phonons was shown to generate a long-lived transient ferroelectric state in the high-symmetry cubic phase $SrTiO_3$ at room temperture[27].

Beyond inversion-symmetry control, circularly polarized THz light offers a pathway to manipulate time-reversal symmetry by generating effective magnetic fields through both off-resonant and resonant phonon-mediated processes[28-35] (Fig. 1a). In the off-resonant regime, the inverse Faraday effect enables light-helicity-dependent ultrafast magnon generation in magnetic materials[36], providing a purely electronic route to light-induced magnetization. In contrast, resonant excitation of axial phonons by circularly polarized THz pulses imparts angular momentum to the lattice, producing an ionic magnetization that can be comparable to electronic contributions[28-31,37,38]. This phonon-driven mechanism was recently confirmed by ultrafast magneto-optic Kerr measurements in $SrTiO_3$ under circularly polarized THz excitation resonant with transverse optical (TO) phonons[32]. Remarkably, the magnetic fields generated by these axial phonons are sufficiently large to reverse the magnetization in adjacent magnetic layers, opening opportunities for ultrafast spintronic devices[28]. Despite these advances, the mechanisms of dynamical multiferroicity resulting from the interplay between inversion and time-reversal symmetry breaking in light-driven phenomena remain largely unexplored.

Here, we demonstrate ultrafast control of time-reversal symmetry in the cubic phase of $SrTiO_3$ (STO) by an off-resonant, elliptically polarized THz pulse. Symmetry-sensitive second harmonic generation (SHG) polarimetry[39-41] reveals time-reversal symmetry breaking under elliptically polarized THz excitation (Fig. 1b). The transient SHG circular dichroism signals obtained under elliptically polarized THz excitation cannot be explained solely by the THz-field-induced SHG (TFISH) from $m3m$ symmetry contribution, indicating an additional emergent SHG contribution. Its azimuthal-angle-dependent polar pattern shows clear mirror symmetry breaking, implying symmetry lowering from cubic $m3m$ to tetragonal $4/mm'm'$. Furthermore, the amplitude of the emergent SHG signal scales linearly with the angular momentum of the elliptically polarized THz pulse, providing compelling evidence for ultrafast time-reversal symmetry breaking in STO. These findings establish a broadly applicable approach for the simultaneous ultrafast control of spatial and time-reversal symmetries, opening new opportunities for dynamical symmetry engineering in quantum materials.

We first introduce a simple and efficient approach for generating elliptically polarized THz pulses using a controlled time delay between two orthogonal $E$-field components[42,43]. As illustrated in Fig. 2a, a vertically polarized THz pulse ($E^{\mathrm{THz}}$) is split by a wire-grid polarizer oriented at +45° into two orthogonal components: the +45° component ($E_{+45}$) is transmitted, while the –45° component ($E_{-45°}$) is reflected. These pulses propagate along a closed triangular loop in a Sagnac configuration. When the two optical paths are equal, the two components recombine after completing the loop into a vertically polarized THz pulse identical to the initial one, without power loss.

Shifting the wire-grid polarizer off-center changes the reflected beam path length, varying the ellipticity of the THz pulse. Figure 2b shows the reflected $E_{-45°}$ (top panel) and the transmitted $E_{+45°}$ (bottom panel) components obtained by electro-optic (EO) sampling at different off-center shifts $\Delta d$ of the wire grid between 0 to 2.5 mm. The transmitted THz pulse is unaffected by the shift, while the reflected THz pulse exhibits a systematic variation of the time delay with $\Delta d$ while maintaining the same waveform. An elliptically polarized THz pulse is obtained by introducing an appropriate time delay between the reflected and transmitted THz pulses (Section S1, Supplementary Information). Figure 2c displays the *x*- and *y*-components ($E_x$ and $E_y$) of elliptically polarized THz pulses with right-handed (RCP, purple) and left-handed (LCP, green) helicities obtained at $\Delta d$ = 2.5 mm and –2.5 mm, respectively. $E_x$ and $E_y$ were calculated from $E_{-45°}$ and $E_{+45°}$ using $E_x = E_{+45°}\cos(45°) + E_{-45°}\cos(-45°)$ and $E_y = E_{+45°}\sin(45°) + E_{-45°}\sin(-45°)$. Notably, the $E_x$ components of the RCP and LCP THz pulses have the same waveforms but opposite signs, while the $E_y$ components are identical. This symmetry between RCP and LCP THz pulse waveforms minimizes artifacts, enabling precise THz-helicity-dependent measurements.

To investigate THz-induced transient symmetry changes, we performed azimuthal angle-dependent SHG polarimetry on STO (001) bulk crystal (*Crystec GmbH*) under both linearly and elliptically polarized THz excitation. We define the azimuthal angle φ for linearly polarized THz pulses by the orientation of the optical plane's intersection with the sample surface relative to the [100] direction of STO. Under THz excitation, the SHG signal $I_{SHG}$ is generated by a crystal with cubic $m3m$ symmetry via a third-order process known as TFISH (Supplementary Note 1). If the inversion symmetry is broken under THz excitation (Supplementary Note 2), additional second-order SHG signals can emerge with dependences on the THz pump and the optical probe polarization distinct from TFISH.

To characterize time-reversal symmetry breaking, we measure SHG circular dichroism $I_{CD} = I_{SHG}(\sigma^+) - I_{SHG}(\sigma^-)$ at $\varphi = 45°$ and a 45° incidence angle under two helicities (RCP and LCP) of elliptically polarized THz pump, as well as two polarities (+45° and –45°) of linearly polarized THz pump (Fig. 3a). We note that the linearly polarized THz pump was obtained by removing the wire-grid polarizer in the Sagnac interferometer. Here, $\sigma^+$ and $\sigma^-$ denote the two circular polarizations of the optical probe. We define THz helicity-dependent SHG dichroism $\Delta I_{CD}$ for elliptically polarized THz pulses as $\Delta CP = I_{CD}(RCP) - I_{CD}(LCP)$, and linear polarization-dependent dichroism as $\Delta LP(45°) = I_{CD}(+45°) - I_{CD}(-45°)$. We additionally measured SHG dichroism $\Delta LP(0°)$ under the same configuration as $\Delta CP$ and $\Delta LP(45°)$, by inserting an additional wire-grid polarizer pulse after the Sagnac interferometer that transmits only the $E_x$ component of the THz pulse. We note that $\Delta LP(0°)$ for $\Delta CP$ and $\Delta LP(45°)$ configuration are different due to their different $E_x$ components. However, to simplify notations we label both quantities $\Delta LP(0°)$, as the specific value is clear from the context in the discussion below.

The emergent SHG signal under elliptically polarized THz excitation can be isolated by comparing these SHG dichroism signals. If the $m3m$ crystal symmetry is unchanged by the THz pulse, the resulting $\Delta CP$ and $\Delta LP(0°)$ are expected to be proportional to $E_x$ (Section S2, Supplementary Information), which is the same for the two configurations, and therefore they would coincide in this case. On the other hand, if time-reversal symmetry is broken by the angular momentum $L_z \sim \boldsymbol{E}\times d\boldsymbol{E}/dt$ of light, the dichroism signal should exhibit an additional component tracking $L_z$. The angular momentum of light vanishes for the LP pulses but is finite for the CP pulses, as shown in Fig. 3b (blue) for the RCP THz pulse obtained at $\Delta d$ = 2.5 mm. Thus, time reversal symmetry breaking should manifest as a difference between ΔCP and ΔLP(0°). On the other hand, LP THz pulses do not break time reversal symmetry, and therefore ΔLP(45°) and the corresponding ΔLP(0°) are expected to coincide.

Figure 3c shows ΔCP (blue), ΔLP(45°) (red), and the corresponding ΔLP(0°) (black) . Both ΔLP(0°) and ΔLP(45°) closely follow the same oscillatory waveform determined by $E_x$, implying that the time reversal symmetry is not violated under linearly polarized THz excitation. On the other hand, ΔCP is dominated by a non-oscillatory peak which is significantly different from ΔLP(0°), as further highlighted by the FFT spectra shown in the inset of Fig. 3c. This demonstrates that additional SHG signal emerges in ΔCP due to breaking of *m*3*m* symmetry. To extract the emergent SHG signal, we plot ΔCP–ΔLP(0°) (blue) and ΔLP(45°)–ΔLP(0°) (red) in Fig. 3d. The emergent SHG signal appears only in the former, and its time dependence closely follows $L_z$, implying that the non-oscillatory SHG signal under elliptically polarized THz excitation results from time-reversal symmetry breaking.

The dependence on azimuthal angle φ further confirms that the emergent SHG signal follows a symmetry distinct from the cubic *m*3*m* group. The dependence of ΔLP(0°) on φ has a maximum φ = 0° and a minimum at 45°, consistent with the pure TFISH following *m*3*m* symmetry (Fig.4a). While ΔCP–ΔLP(0°) also shows a four-fold rotation symmetry, it exhibits a maximum at φ = 60° and a minimum at 105°, which is not consistent with the mirror symmetries of the *m*3*m* group (Fig.4b). The φ-dependent polar patterns further highlight the *m*3*m* symmetry breaking by the elliptically polarized THz pulses, as shown in Figs. 4c and 4d. While the ΔLP(0°) polar pattern is aligned with the [100] and [010] axes, the polar pattern for ΔCP–ΔLP(0°) is rotated, breaking the mirror symmetries of the *m*3*m* group. According to Neumann's principle, this must result from the broken cubic symmetry of the crystal.

The mirror-symmetry breaking in the SHG polar pattern under elliptically polarized THz excitation reflects time-reversal symmetry breaking in STO driven by a finite $L_z$, as follows. In cubic *m*3*m* symmetry, there are four mirror planes perpendicular to the [100], [110], [010], and $[1\bar{1}0]$ directions and a four-fold rotation axis along the *z*-direction (Fig. 4e). Therefore, the polar pattern of TFISH should exhibit a four-fold anisotropy symmetric with respect to the principal axes, as observed in ΔLP(0°). When $L_z$ is finite, the four-fold rotational symmetry is retained (Fig. 4f) but all in-plane mirror symmetries are broken and replaced by time-reversal mirrors (red dotted lines), resulting in symmetry lowering from cubic *m*3*m* to tetragonal 4/*mm*′*m*′ (Supplementary Note 2). The polar pattern for ΔCP–ΔLP(0°) clearly violates the mirror symmetries, consistent with symmetry lowering to 4/*mm*′*m*′. By considering the third-order *m*3*m* *i*-tensors (red line) and 4/*mm*′*m*′ *c*-tensors (black line), which are even and odd under time-reversal, respectively, we fitted the polar patterns for ΔLP(0°) and ΔCP–ΔLP(0°) (Figs. 4c and 4d) (Supplementary Note 3). The fits are in excellent agreement with the experimental results, supporting the proposed mechanism of symmetry breaking under elliptically polarized THz excitation.

We quantitatively tested the effects of THz ellipticity by varying the offset Δ*d* in the off-centered Sagnac interferometer. Figures 5a and 5b show that $L_z$ and ΔCP–ΔLP(0°) both depend on Δ*d*. At Δ*d* = 0 mm, the THz pulse is linearly polarized and $L_z$ vanishes. Both $L_z$ and ΔCP–ΔLP(0°) progressively increase with increasing Δ*d*, indicating a correlation between ΔCP and $L_z$. As shown in Fig. 5c, ΔCP–ΔLP(0°) depends linearly on $L_z$, demonstrating that time-reversal symmetry in STO can be optically controlled by the ellipticity of THz excitation.

Figure 5d illustrates the THz-induced transient time-reversal symmetry breaking in STO, resulting in the symmetry lowering from cubic to tetragonal. An elliptically polarized THz pulse can break time-reversal symmetry either by inducing electronic magnetization via inverse Faraday effect[33-35] or by inducing ionic magnetization via dynamical multiferroicity[28,29,31,32]. The latter mechanism can be also interpreted as ionic inverse Faraday effect[33]. Since STO is a wide-bandgap insulator with a quenched spin and orbital electronic state, the former contribution is expected to be small. On the other hand, since the frequency of TO phonons (3 THz at room temperature)[44] is much higher than that of the THz pulse (0.5 THz) used in our experiments, one might also expect the ionic contribution from the population of TO phonons to be small. It is therefore notable that we observe a significant THz-helicity-dependent magneto-optic signal in STO, indicating efficient generation of magnetic moments even under off-resonant conditions (Section S3, Supplementary Information).

These observations motivate a closer examination of the intrinsic properties of STO that enable magnetic-moment generation under non-resonant circularly polarized excitation. At cryogenic temperatures, this material forms a quantum paraelectric (QPE) state[45] characterized by finite-displacement ionic energy minima whose ferroelectric ordering is prevented by quantum effects. In the dielectric state at room temperature, the shallow energy landscape results in a large position uncertainty, which invalidates the classical polarization dynamics models. Motivated by this observation, we developed a minimal quantum model of dynamical ionic states, which elucidates a large role of incipient ferroelectric displacement potential in time-reversal symmetry breaking even at room temperature (Section S4, Supplementary Information), as outlined below.

In the harmonic approximation for the quantum Ti ion dynamics relative to the coordinating oxygen octahedron, the Hamiltonian describing ionic dynamics in the *xy* plane is

$$H = -\frac{\hbar^2\nabla^2}{2M} + \frac{M\omega_{TO}^2\boldsymbol{r}^2}{2}, \qquad (1)$$

where $M$ is the effective mass of TO mode, and the mode displacement $\boldsymbol{u} = \boldsymbol{r}$. By solving Eq. 1, the ground-state ionic wavefunction $\psi_0$ can be obtained, which is isotropic and does not carry polarization or magnetization. The two lowest-energy excited states $\psi_x$, $\psi_y$ are two-fold degenerate due to the four-fold in-plane rotation symmetry. Their coherent superpositions $\psi_{L,R} = \frac{1}{\sqrt{2}}(\psi_x \pm i\psi_y)$ exhibit an isotropic amplitude distribution similar to that of the ground state, while their phases vary azimuthally with the polar angle in opposite directions. Therefore, these states carry circulating ionic currents generating ionic magnetic moments that break time-reversal symmetry, and can be selectively excited by LCP and RCP light, respectively.

In thermal equilibrium, $\psi_x$ and $\psi_y$ are equally populated and lack mutual phase coherence, so the equilibrium state does not carry net polarization or magnetic moment. A key insight is that THz-driven polarization appears only in superpositions of these states with $\psi_0$, while $\psi_{L,R}$ alone carry circulating ionic currents which cannot be described in terms of classical dynamical polarization. Consequently, the classical formula $\boldsymbol{m_c} \propto \boldsymbol{p} \times \dot{\boldsymbol{p}}$ can drastically underestimate the ionic contribution to time-reversal symmetry breaking by circularly polarized light.

We confirmed this insight by the simulations of the density matrix dynamics based on the Lindblad equation (Fig. S10, Supplementary Information). The simulations confirm that non-resonant LCP (RCP) THz excitation of an initially incoherent room-temperature equilibrium state generates coherence between $\psi_x$ and $\psi_y$ in, producing transient non-oscillatory ionic magnetization $m_q$ which is almost an order of magnitude larger than would be expected from classical ionic motion. This non-classical magnetic moment enhancement originates from the shallow displacement potential enabling circulating ionic states with zero expectation value of displacement and a large displacement uncertainty. In contrast, in dielectric materials characterized by a narrow centrosymmetric displacement potential, the ionic position is well-defined, and the pump-driven states are well-described by the classical polarization dynamics.

In summary, we have experimentally demonstrated ultrafast time-reversal symmetry breaking in a non-magnetic STO at room temperature due to elliptically polarized THz excitation. The central role of THz helicity is confirmed by comparison with the transient SHG circular dichroism signals obtained under linearly polarized THz pulses, which exhibit a qualitatively different symmetry and time dependence. We extracted the emergent SHG signals under elliptically polarized THz excitation and found that its azimuth-angle-dependent SHG polar pattern exhibits a clear breaking of mirror symmetry. These transient SHG signals lacking mirror symmetry imply symmetry-lowering from cubic $m3m$ to magnetic tetragonal $4/mm'm'$ due to ultrafast time-reversal symmetry breaking by elliptically polarized THz excitation. We observe a linear relationship between the emergent SHG signals and the angular momentum of the THz field, confirming the key role of angular momentum in controlling the time-reversal symmetry breaking in STO. Our findings elucidate the light-matter interactions that give rise to symmetry reduction under circularly polarized light and identify a straightforward approach to controlling time-reversal symmetry on ultrafast timescales, opening new opportunities for ultrafast lattice and electronic structure engineering.

## Methods

### Characterization of THz electric field

We used a femtosecond amplified laser with 1 KHz repetition rate and the pulse energy of 4 mJ (Astrella, *Coherent*) to generate single-cycle THz pulses using $LiNbO_3$ prism with the tilted-pulse-front technique. The THz power and beam size were measured using a THz power meter and THz camera. The calculated peak intensity is ~300 KV/cm with the center frequency of 0.5 THz. The initial THz polarization was defined by a vertically aligned wire-grid polarizer, and elliptically polarized THz pulses were produced by an off-centered Sagnac interferometer. To produce linearly polarized THz pulses, an additional wire-grid polarizer set along 45° or 0° was inserted after the off-centered Sagnac interferometer. The time profile of the THz pulse was characterized by electro-optic sampling using a 280-μm-thick GaP (110) crystal.

### Second-harmonic generation circular dichroism

To obtain second-harmonic generation (SHG) circular dichroism signals, we produced circularly polarized optical probe pulses (800 nm) with two helicities ($\sigma^+$ and $\sigma^-$) using a zeroth-order quarter-wave-plate (QWP). The circular polarization states were precisely characterized by polarizer-QWP-analyzer polarimetry. To eliminate any second-harmonic light induced by transparent optics, a long-pass-filter with the cut-off frequency of 600 nm was inserted before the sample. The probe pulse was tightly focused on the sample surface at the 45° incidence angle using a convex lens with a focal length of 150 mm. We measured the SHG signals using a photomultiplier tube with a current preamplifier. The fundamental light was blocked by short-pass and band-pass filter to detect only the second-harmonic light (400 nm) induced by the sample. An analyzer oriented parallel to the $SrTiO_3$ [010] axis was used to detect the SHG output. Azimuthal angle-dependent SHG circular dichroism signals were obtained by rotating the sample from 0° to 180°, and duplicated for the range from 180° to 360° based on the rotational symmetry.

## Acknowledgement

I.H.C. was supported by the National Research Foundation of Korea (NRF) grant funded by the Korea government (MSIT) (No. RS-2024-00351794). I.H.C. and K.A.N. acknowledge support from Office of Naval Research Grant No. 14035670. S.U. was supported by the US National Science Foundation (NSF) award No. ECCS- 2448290. The work at UMN (S.V., S.G.J, and B.J.) was primarily supported from the Air Force Office of Scientific Research Multi University Research Initiative (AFOSR MURI, Award No. FA9550-25-1-0262). This work was supported by the Hankuk University of Foreign Studies Research Fund.

## Contributions

I.H.C conceived the idea and designed the experiments. I.H.C performed SHG polarimetry and symmetry analysis. S.U. performed numerical simulations. S.V., S.G.J., and B.J. prepared the sample. I.H.C. and K.A.N. drafted the manuscript initially and revised it based on input and feedback from all the authors. K.A.N. directed this project.

## Corresponding author

Correspondence to Keith A. Nelson.

## Competing interests

The authors declare no competing interests.

## Data availability

All data supporting the results within this paper and supporting information are available from the corresponding author upon reasonable request. Source data are provided with this paper.

## Code availability

All numerical simulation codes employed this work are available from the corresponding author upon reasonable request.

## Figures

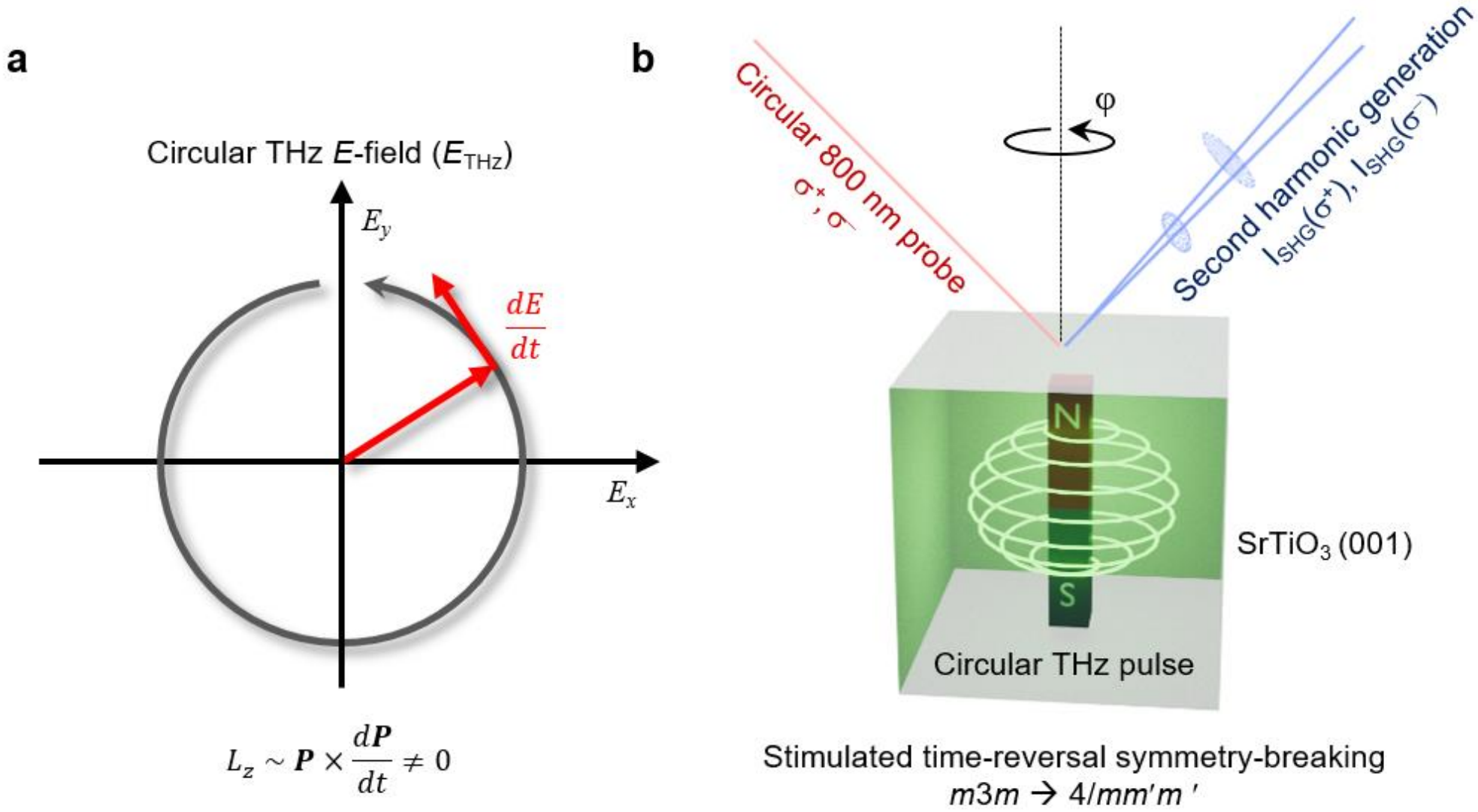

**Figure 1. Time-reversal symmetry breaking by elliptically polarized THz pump. (a)** Circularly polarized THz *E*-field ($E_{THz}$) in the transverse plane. The *x*- ($E_x$) and *y*-components ($E_y$) of $E_{THz}$ induce a rotating spontaneous polarization $\boldsymbol{P}$ carrying a non-zero angular momentum $L_z \propto \boldsymbol{P} \times \mathrm{d}\boldsymbol{P}/\mathrm{d}t$. **(b)** Schematic illustration of second-harmonic generation (SHG) circular dichroism polarimetry of $SrTiO_3$ under circularly polarized THz excitation. Emergent SHG signals induced by symmetry lowering are distinguished from THz-field-induced SHG signals using two different helicities of circularly polarized optical probe ($\sigma^+$, and $\sigma^-$) and THz pump (right-handed, and left-handed). We obtained SHG signals $I_{SHG}(\sigma^+, \sigma^-)$ at the incidence angle of 45°, enabling heterodyne detection of SHG *E*-field components by interference with the background SHG signals.

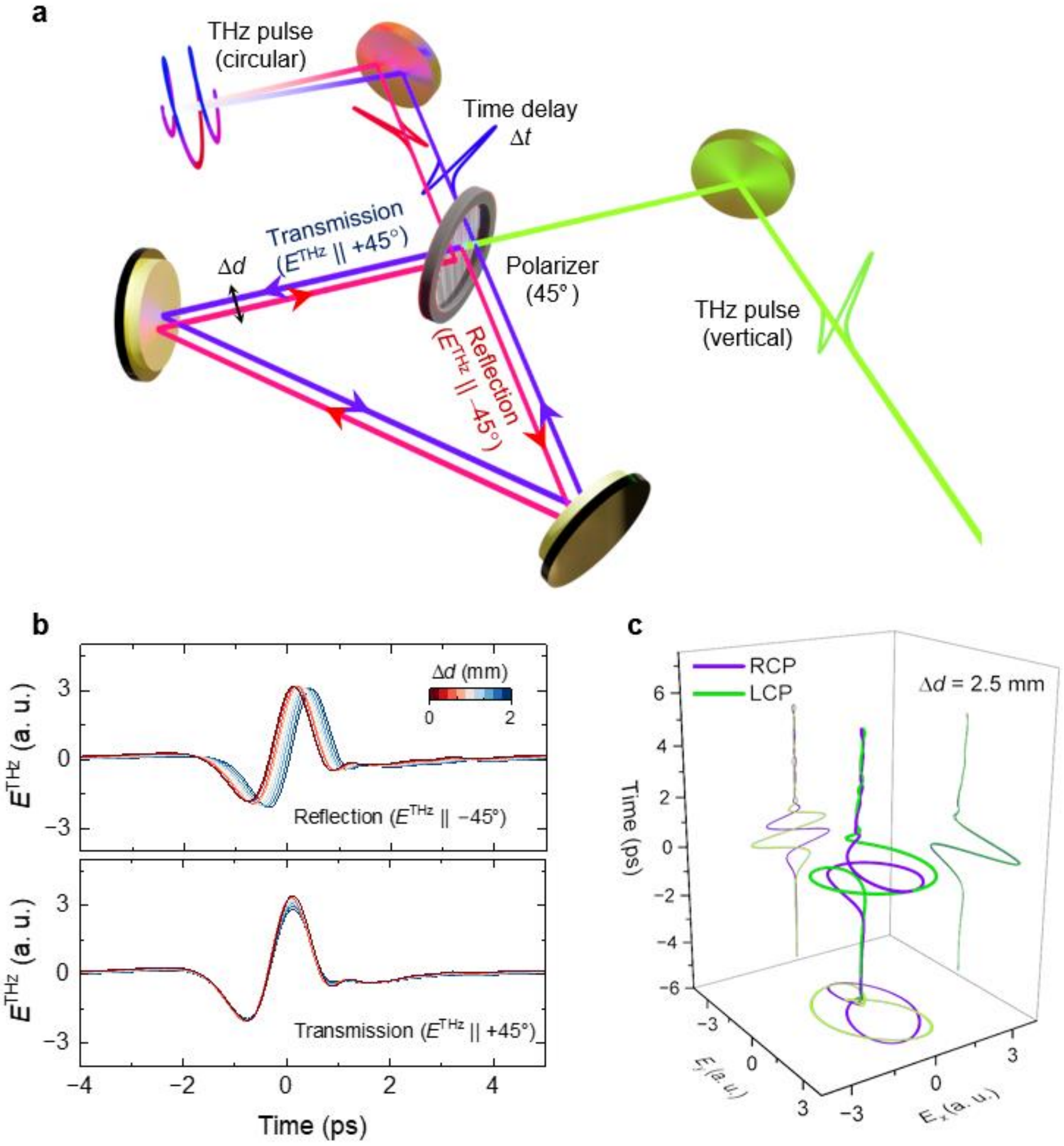

**Figure 2. Generation of two symmetric opposite helicities of elliptically polarized THz pulses using an off-centered Sagnac interferometer configuration. (a)** Schematic illustration of an off-centered Sagnac interferometer. The vertically polarized THz pulse is split into two orthogonal components, polarized at +45° (transmission) and −45° (reflection) using a wire-grid polarizer. The two pulses propagate around a closed loop and recombine without energy loss. A time delay between the two pulses is introduced by adjusting the position of the wire-grid polarizer. **(b)** THz time-profile for the pulses reflected (top) and transmitted (bottom) from the wire-grid polarizer, obtained by electro-optic sampling. The time delay of the reflected pulse is controlled by the off-center position Δ*d* of the wire-grid polarizer. **(c)** 3d plot of right-handed (RCP, purple) and left-handed (LCP) elliptical THz pulses, and their projection on the *xy* plane. We obtained two helicities of THz pulses by rotating the wire-grid polarizer in the Sagnac interferometer to ±45°. Symmetric opposite elliptical polarizations are obtained by opposite off-center displacements of the Sagnac interferometer.

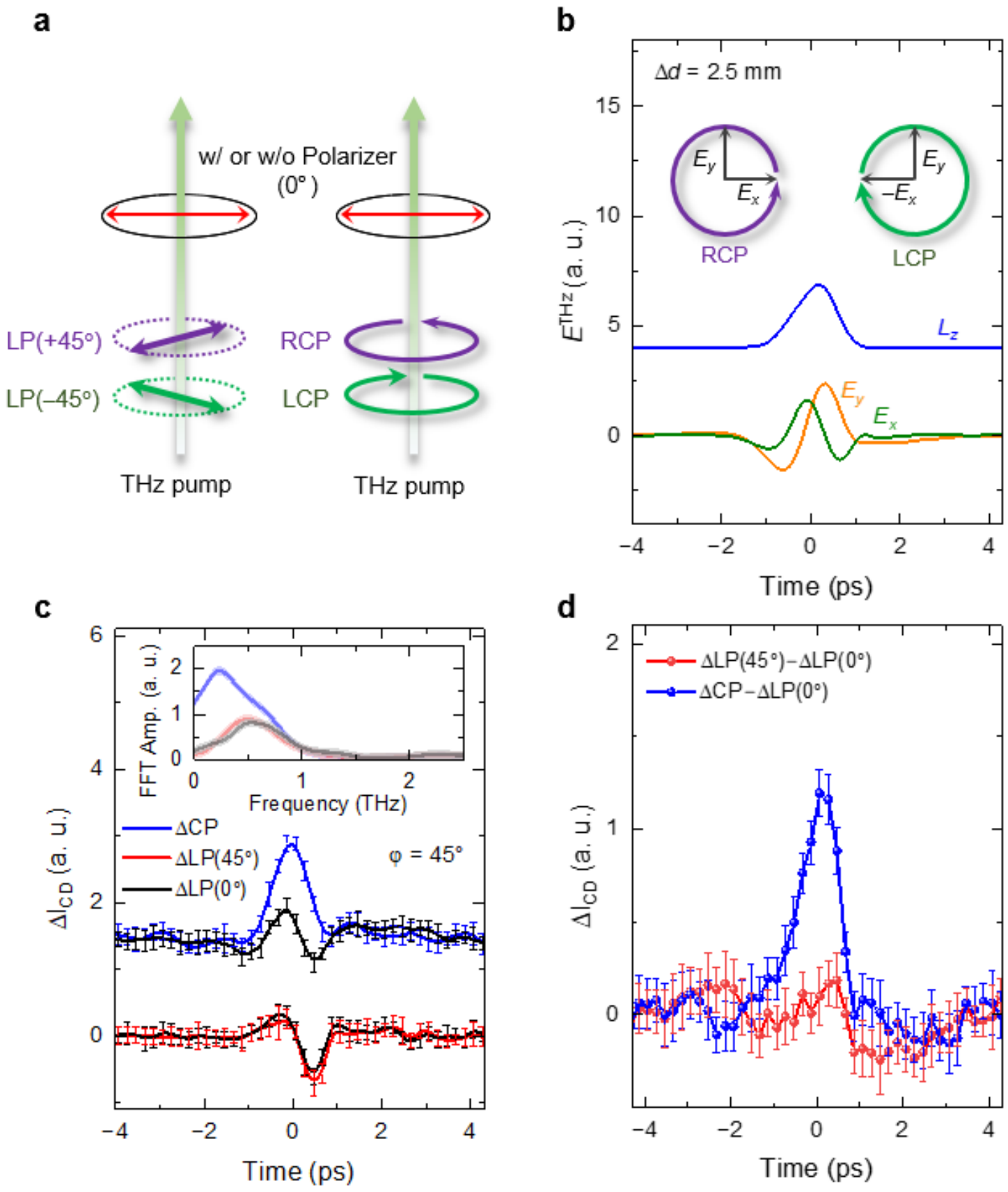

**Figure 3. Transient SHG circular dichroism polarimetry under different polarizations of THz excitation. (a)** Linearly and elliptically polarized THz pulses obtained with and without a second wire-grid polarizer, respectively. When alternating the helicities of THz pulses, only the $E_x$-components change sign while the waveform remains the same. **(b)** THz time-profiles of $E_x$ (green), $E_y$ (orange), and $L_z$ (blue) obtained by electro-optic sampling at $\Delta d = 2.5$ mm (RCP). **(c)** Transient SHG circular dichroism signals $\Delta I_{CD}$ under linearly (ΔLP) and elliptically (ΔCP) polarized THz excitation, obtained by subtracting the responses for two opposite THz polarities and helicities, respectively, at sample azimuth angle φ = 45°. Inset shows the corresponding FFT spectra. **(d)** Emergent SHG signals for linearly and elliptically polarized THz excitation, calculated as ΔLP(45°)–ΔLP(0°) (red) and ΔCP−ΔLP(0°) (blue), respectively. We note that ΔLP(0°) originates solely from the THz-field-induced SHG (TFISH) process expected for *m*3*m* symmetry.

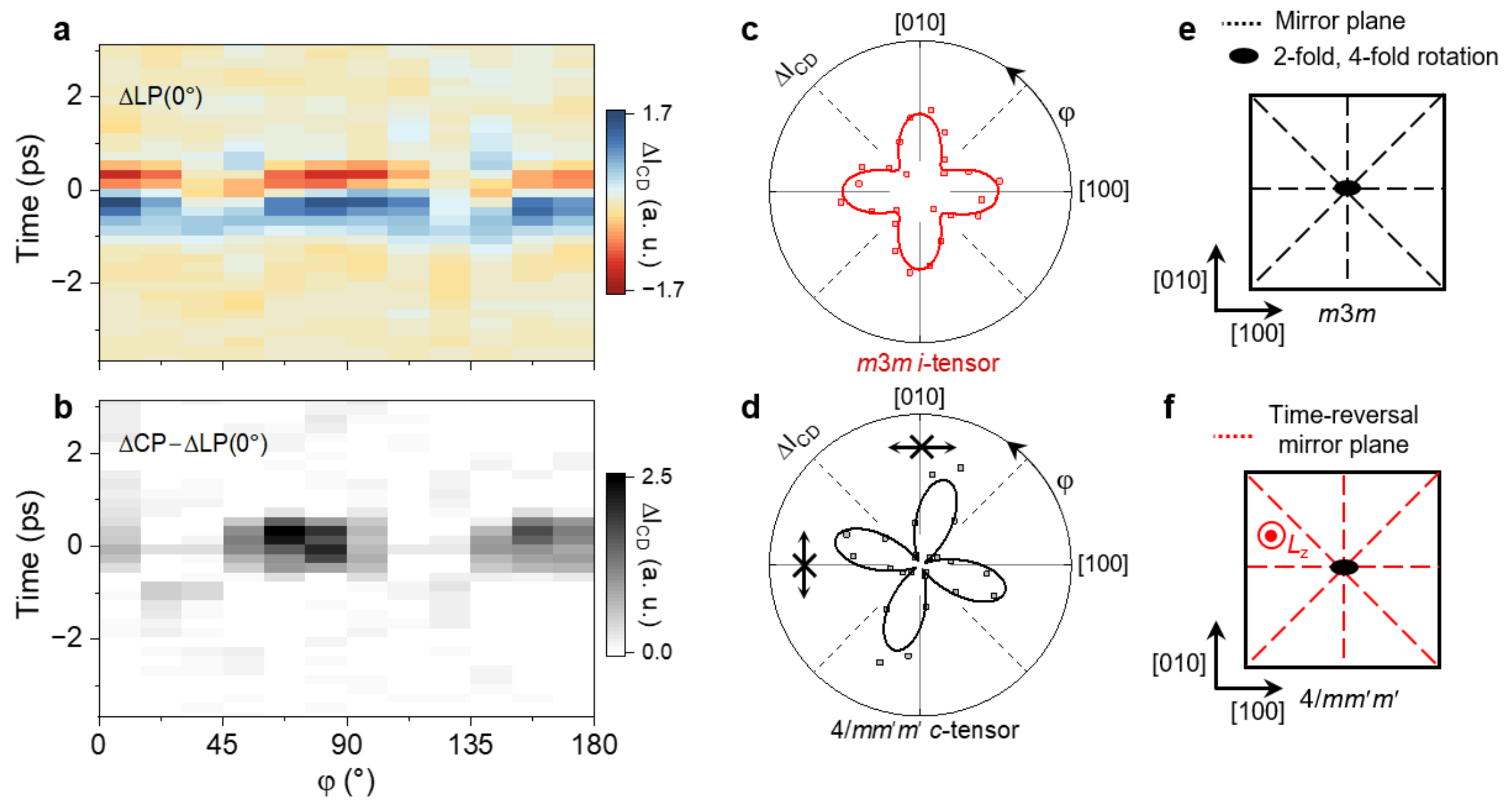

**Figure 4. Dependence of $\Delta I_{SHG}$ on the azimuth angle φ under linearly and elliptically polarized THz excitation. (a, b)** φ-dependence of ΔLP(0°) originating from TFISH **(a)** and of ΔCP−ΔLP(0°) originating from emergent SHG due to spatial symmetry lowering **(b)**. φ = 0° corresponds to the orientation of analyzer parallel to the [100] axis of $SrTiO_3$. **(c, d)** φ-dependent $\Delta I_{CD}$ polar patterns for ΔLP(0°) **(c)** and ΔCP−ΔLP(0°) **(d)**. Solid lines represent a fitting curve considering *m*3*m* *i*-tensor (red) and 4/*mm′m′* (black) *c*-tensor. **(e, f)** Illustration of mirror plane (black dashed lines), time-reversal mirror plane (red dotted lines) and two- and four-fold rotation axis (oval symbols) for **(e)** *m*3*m* symmetry and **(f)** 4/*mm′m′* symmetry.

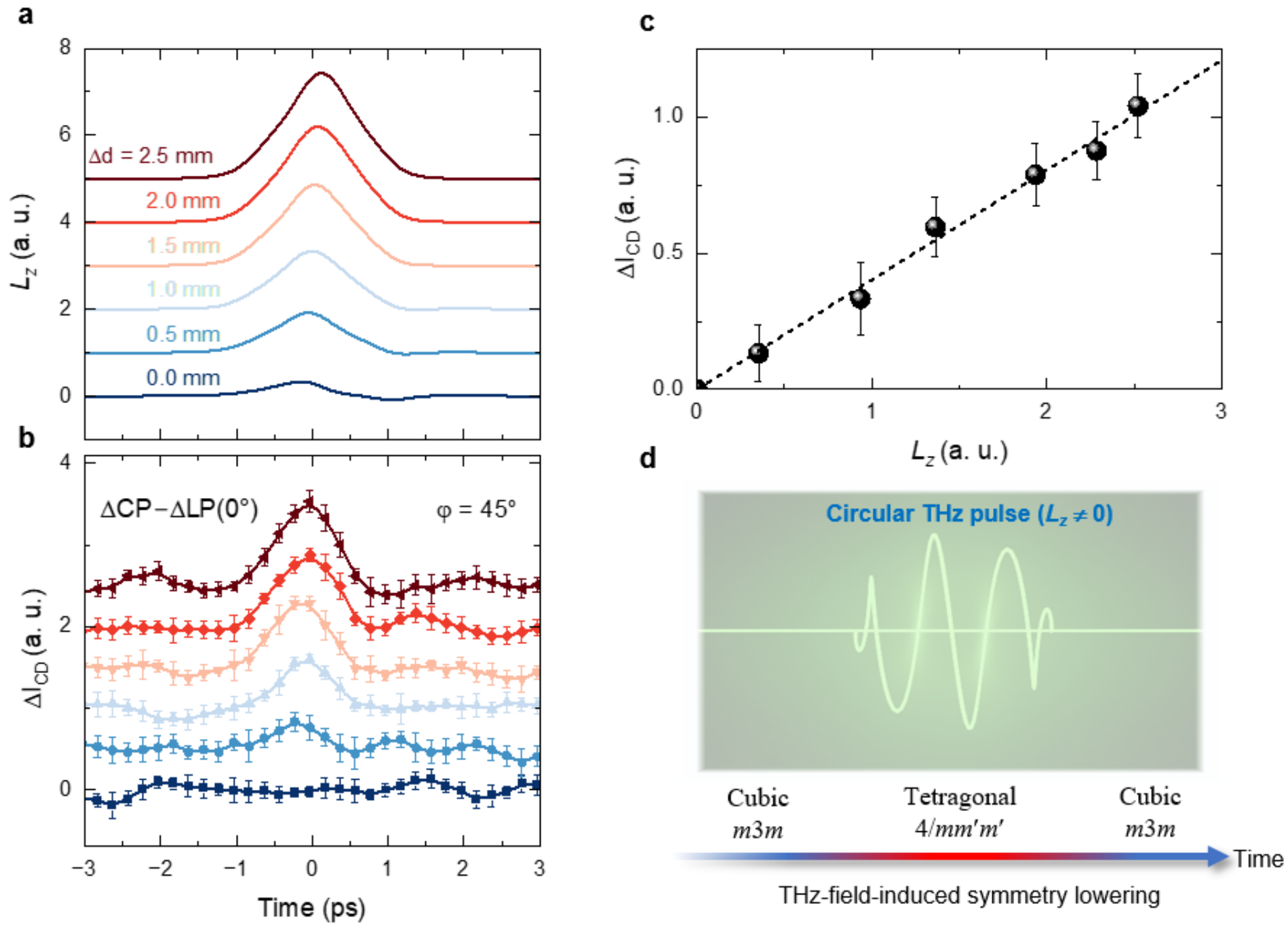

**Figure 5. Correlation between the angular momentum of the THz *E*-field and emergent SHG signals from time-reversal symmetry breaking. (a)** Angular momentum $L_z$ determined from the time profiles of the THz *E*-field, at the labeled Δ*d*. **(b)** Pure emergent SHG signals ΔCP−ΔLP(0°) obtained at φ = 45°, at the same values of Δ*d* as in (a). **(c)** Peak intensity of ΔCP−ΔLP(0°) as a function of $L_z$. The dashed line represents a linear fitting curve. **(d)** Illustration of ultrafast time-reversal symmetry breaking under elliptically polarized THz excitation, accompanied by symmetry lowering from cubic *m*3*m* to tetragonal magnetic 4/*mm*′*m*′.

# Supplementary Information

## Light-Driven Ultrafast Control of Time-Reversal Symmetry in $SrTiO_3$

In Hyeok Choi[1,†], Sergei Urazhdin[2], Shivasheesh Varshney[3], Seung Gyo Jeong[3,4], Bharat Jalan[3] and Keith A. Nelson[1,*]

[1]*Department of Chemistry, Massachusetts Institute of Technology, Cambridge, Massachusetts 02139, United States*

[2]*Department of Physics, Emory University, Atlanta, Georgia 30322, United States*

[3]*Department of Chemical Engineering and Materials Science, University of Minnesota−Twin Cities, Minneapolis, Minnesota 55455, United States*

[4]*Department of Physics, Hankuk University of Foreign Studies, Yongin 17035, Republic of Korea*

[†] First author

[*]Corresponding authors: kanelson@mit.edu

**Supplementary Note 1 – Analytic solution for THz-field-induced SHG**

Under THz excitation, THz pump and optical probe fields induce second-harmonic light, following, $E_j(2\omega) \sim \chi^i_{jklm} E_j^{THz} E_k^p(\omega) E_l^p(\omega)$, namely THz-field-induced second-harmonic generation (TFISH). Here, $E^{\mathrm{THz}}$ and $E^p$ are the of THz pump and optical probe electric fields, respectively, and $\chi^i$ is third-order susceptibility tensors for cubic *m*3*m* symmetry. The TFISH signals are sensitively modulated under the polarization states of both THz and optical fields, facilitating the investigation of symmetry. For SHG polarimetry measurements, we employed circularly polarized optical pulses, which can provide phase information about susceptibility tensor components. *E*-field components of the circularly polarized optical pulses for two helicities ($\sigma^+$ and $\sigma^-$) can be written as $\sigma^{+,-} = E_x^p \pm iE_y^p$. We obtained TFISH signals at the incidence angle of 45°, where non-zero background SHG signals $E_0$ can be induced by *m*3*m* bulk electric quadrupole and 4*mm* surface electric dipole[1,2] contributions.

Table S1 summarizes the analytic solution of complex *S*-polarized THz-field-induced second-harmonic electric field for $\sigma^+$ and $\sigma^-$ as a function of azimuth angle φ. The *S*-polarized second-harmonic electric field Fresnel coefficients are given constant real values for $\sigma^+$ and $\sigma^-$ due to negligible extinction coefficients at 800 nm and 400 nm[3]. When we measure second-harmonic light through a detector, the intensity is obtained ($I_{\mathrm{SHG}} \sim |E(2\omega)|^2$), resulting in both TFISH intensity ($|E_{x,y}|^2$) and heterodyne components ($E_0E_{x,y}$). Both real and imaginary parts show four-fold rotational symmetry in their polar patterns due to cubic *m*3*m* symmetry. We note that the only difference between $\sigma^+$ and $\sigma^-$ lies in the sign of the imaginary components of the electric field, while the real part remains unchanged. This allows us to extract pure heterodyne signals in SHG circular dichroism signals $I_{CD}$ by subtracting $I_{SHG}(\sigma^+)$ and $I_{SHG}(\sigma^-)$, eliminating all intensity contributions. On the other hand, we can selectively obtain the $E_x$ and $E_y$ components of THz field by adding (RCP + LCP) and subtracting (RCP – LCP) $I_{SHG}$ obtained under right-handed (RCP) and left-handed (LCP) THz excitation. Because the two THz helicities produce the same $E_y$ component but opposite signs in the $E_x$ component, RCP + LCP contains only the intensity term and the $E_y$-related heterodyne contribution, whereas the RCP – LCP isolates the $E_x$-related heterodyne contribution. Table S2 shows the analytic solution for various combinations of THz pump and optical probe polarization states for φ = 0° and 45°. We note that only the lowest (second-order) terms were considered. We expected that $I_{CD}$(RCP) – $I_{CD}$(LCP) should exhibit only $E_x$-related heterodyne signals if there is only a TFISH contribution under elliptically polarized THz excitation.

**Table S1. φ-dependent THz-field-induced second harmonic electric field with $\sigma^+$, and $\sigma^-$ probe polarization for cubic *m*3*m*.**

| Optical probe polarization | Second harmonic electric field | |
|---|---|---|
| | Real | Imaginary |
| $\sigma^+$ | $E_0$<br>$+(\chi^i_{xxxx} + \chi^i_{yyzz} - \chi^i_{yzyz} + \chi^i_{yzzy})E_y$<br>$+(\chi^i_{xxxx} - \chi^i_{yyzz} - \chi^i_{yzyz} - \chi^i_{yzzy})E_y \cos(4\varphi)$<br>$-(\chi^i_{xxxx} - \chi^i_{yyzz} - \chi^i_{yzyz} - \chi^i_{yzzy})E_x \sin(4\varphi)$ | $E_0$<br>$-(\chi^i_{xxxx} + \chi^i_{yyzz} - \chi^i_{yzyz} + \chi^i_{yzzy})E_x$<br>$+(\chi^i_{xxxx} - \chi^i_{yyzz} - \chi^i_{yzyz} - \chi^i_{yzzy})E_x \cos(4\varphi)$<br>$+(\chi^i_{xxxx} - \chi^i_{yyzz} - \chi^i_{yzyz} - \chi^i_{yzzy})E_y \sin(4\varphi)$ |
| $\sigma^-$ | $E_0$<br>$+(\chi^i_{xxxx} + \chi^i_{yyzz} - \chi^i_{yzyz} + \chi^i_{yzzy})E_y$<br>$+(\chi^i_{xxxx} - \chi^i_{yyzz} - \chi^i_{yzyz} - \chi^i_{yzzy})E_y \cos(4\varphi)$<br>$-(\chi^i_{xxxx} - \chi^i_{yyzz} - \chi^i_{yzyz} - \chi^i_{yzzy})E_x \sin(4\varphi)$ | $E_0$<br>$+(\chi^i_{xxxx} + \chi^i_{yyzz} - \chi^i_{yzyz} + \chi^i_{yzzy})E_x$<br>$-(\chi^i_{xxxx} - \chi^i_{yyzz} - \chi^i_{yzyz} - \chi^i_{yzzy})E_x \cos(4\varphi)$<br>$-(\chi^i_{xxxx} - \chi^i_{yyzz} - \chi^i_{yzyz} - \chi^i_{yzzy})E_y \sin(4\varphi)$ |

**Table S2. THz-field-induced SHG signals for several polarization states of THz pump and optical probe pulse at φ = 0° and 45°. The $E_x$ and $E_y$ components of THz pulse are highlighted by blue and red color, respectively.**

| Azimuth angle | Optical probe polarization | THz pump polarization | w/o polarizer | With polarizer | |
|---|---|---|---|---|---|
| | | | | +45° | 0° |
| φ = 0° | $\sigma^+ + \sigma^-$ | RCP + LCP | $\lvert E_x\rvert^2 + \lvert E_y\rvert^2 + E_0E_y$ | $\lvert E_x\rvert^2 + \lvert E_y\rvert^2 + E_0E_y$ | $\lvert E_x\rvert^2$ |
| | | RCP – LCP | 0 | 0 | 0 |
| | $\sigma^+ - \sigma^-$ | RCP + LCP | 0 | 0 | 0 |
| | | RCP – LCP | $E_0E_x$ | $E_0E_x$ | $E_0E_x$ |
| φ = 45° | $\sigma^+ + \sigma^-$ | RCP + LCP | $\lvert E_x\rvert^2 + \lvert E_y\rvert^2 + E_0E_y$ | $\lvert E_x\rvert^2 + \lvert E_y\rvert^2 + E_0E_y$ | $\lvert E_x\rvert^2$ |
| | | RCP – LCP | 0 | 0 | 0 |
| | $\sigma^+ - \sigma^-$ | RCP + LCP | $E_0E_y$ | $E_0E_y$ | 0 |
| | | RCP – LCP | $E_0E_x$ | $E_0E_x$ | $E_0E_x$ |

## Supplementary Note 2 – Symmetry lowering of cubic *m*3*m* with a magnetic moment

Centrosymmetric cubic *m*3*m* is the highest crystalline symmetry, exhibiting 13 rotation axes, 9 mirror planes, and one inversion center. If there is an effective magnetic moment along the [001]-axis and it couples to the lattice system, then time-reversal symmetry breaking should lead to symmetry lowering. When a magnetic moment is parallel to a mirror plane, its mirror image reverses sign and restores the original state after applying the time-reversal symmetry operator. This is valid for (100), (010), (110), and ($1\bar{1}0$) planes, and corresponding mirror symmetry operators $m$ should be changed into $m'$, where the superscript represents a time-reversal symmetry operator. On the other hand, a mirror plane (001), perpendicular to the magnetic moment, doesn't change the mirror image of magnetic moment, implying the conservation of mirror symmetry. However, for five mirror planes (011), (101), ($\bar{1}11$), ($1\bar{1}1$), and ($11\bar{1}$), the magnetic moment does not return to its original orientation even after applying time-reversal symmetry, leading to mirror-symmetry breaking. Furthermore, vertices of three-fold rotation axes, and four- and two-fold rotation axes that are not parallel to the magnetic moment, should be broken, resulting in the symmetry lowering into tetragonal $4/mm'm'$. In this magnetic point group, time-reversal-symmetry-breaking-induced SHG can arise, producing SHG polarization patterns distinct from those of TFISH.

**Supplementary Note 3 – Analytic solution for SHG induced by time-reversal symmetry breaking**

In $4/mm'm'$ symmetry, time invariant ($i$-tensor) and non-invariant ($c$-tensor) second-order susceptibility tensor components can contribute SHG signals, which are even and odd respectively under time-reversal. Because $4/mm'm'$ has inversion symmetry, we considered an electric quadrupole (EQ) contribution[4] to obtain the SHG analytic solution. Furthermore, we also accounted the surface electric dipole (ED) contribution from $4m'm'$ that is the surface symmetry of $4/mm'm'$. All non-zero susceptibility tensor components were found based on symmetry analysis with generating matrixes, introduced in *Briss* (1964)[5]. The expected SHG intensity was obtained from the square of the coherent sum of all second-harmonic electric field contributions $I_{SHG} \sim |\sum E(2\omega)|^2$.

1. *Bulk electric quadrupole*

SHG signals can be induced via the EQ process even in centrosymmetric materials, described by $E_j(2\omega) = \chi^{i,c}_{jklm} E_k \nabla_l E_m$. Here, $\chi^{i,c}$ are time invariant and non-invariant third-order susceptibility tensors, respectively. To preserve the time-reversal symmetry of second-harmonic light, time non-invariant tensors are pure imaginary, whereas time invariant tensors are pure real[6]. We calculated the SHG response from the EQ process analytically for the $4/mm'm'$ point group (Tables S3 and S4).

**Table S3. φ-dependent THz-field-induced second harmonic electric field with $\sigma^+$, and $\sigma^-$ probe polarization for $4/mm'm'$ ($i$-tensor).**

| Optical probe polarization | Second harmonic electric field | |
|---|---|---|
| | Real | Imaginary |
| $\sigma^+$ | $(\chi^i_{xxxx} - \chi^i_{xxyy} - \chi^i_{xyxy} - \chi^i_{xyyx})\sin(4\varphi)$ | $(\chi^i_{xxxx} + \chi^i_{xxyy} + \chi^i_{xxzz} - \chi^i_{xyxy} + \chi^i_{xyyx} + \chi^i_{xzzx}) - (\chi^i_{xxxx} - \chi^i_{xxyy} - \chi^i_{xyxy} - \chi^i_{xyyx})\cos(4\varphi)$ |
| $\sigma^-$ | $(\chi^i_{xxxx} - \chi^i_{xxyy} - \chi^i_{xyxy} - \chi^i_{xyyx})\sin(4\varphi)$ | $-(\chi^i_{xxxx} + \chi^i_{xxyy} + \chi^i_{xxzz} - \chi^i_{xyxy} + \chi^i_{xyyx} + \chi^i_{xzzx}) + (\chi^i_{xxxx} - \chi^i_{xxyy} - \chi^i_{xyxy} - \chi^i_{xyyx})\cos(4\varphi)$ |

**Table S4. φ-dependent THz-field-induced second harmonic electric field with $\sigma^+$, and $\sigma^-$ probe polarization for $4/mm'm'$ ($c$-tensor).**

| Optical probe polarization | Second harmonic electric field | |
|---|---|---|
| | Real | Imaginary |
| $\sigma^+$ | $(\chi^c_{xxxx} + \chi^c_{xxyy} + \chi^c_{xxzz} - \chi^c_{xyxy} + \chi^c_{xyyx} + \chi^c_{xzzx}) - (\chi^c_{xxxx} - \chi^c_{xxyy} - \chi^c_{xyxy} - \chi^c_{xyyx})\cos(4\varphi)$ | $(\chi^c_{xxxx} - \chi^c_{xxyy} - \chi^c_{xyxy} - \chi^c_{xyyx})\sin(4\varphi)$ |
| $\sigma^-$ | $-(\chi^c_{xxxx} + \chi^c_{xxyy} + \chi^c_{xxzz} - \chi^c_{xyxy} + \chi^c_{xyyx} + \chi^c_{xzzx}) + (\chi^c_{xxxx} - \chi^c_{xxyy} - \chi^c_{xyxy} - \chi^c_{xyyx})\cos(4\varphi)$ | $(\chi^c_{xxxx} - \chi^c_{xxyy} - \chi^c_{xyxy} - \chi^c_{xyyx})\sin(4\varphi)$ |

2. *Surface electric dipole*

At the surface, out-of-plane mirror symmetry is broken, resulting in symmetry lowering into $4m'm'$. In this case, there are non-zero second-order susceptibility tensors, and SHG can be induced by ED process, with $E_j(2\omega) = \chi^{i,c}_{jkl} E_k E_l$ (Tables S5 and S6). Here, $\chi^{i,c}$ are time-invariant and non-invariant second-order susceptibility tensors. We note that both *i*- and *c*-tensors are independent to φ.

**Table S5. φ-dependent THz-field-induced second harmonic electric field with $\sigma^+$, and $\sigma^-$ probe polarization for $4m'm'$ (*i*-tensor).**

| Optical probe polarization | Second harmonic electric field | |
|---|---|---|
| | Real | Imaginary |
| $\sigma^+$ | 0 | $\chi^i_{xxz} + \chi^i_{xzx}$ |
| $\sigma^-$ | 0 | $-\chi^i_{xxz} - \chi^i_{xzx}$ |

**Table S6. φ-dependent THz-field-induced second harmonic electric field with $\sigma^+$, and $\sigma^-$ probe polarization for $4m'm'$ (*c*-tensor).**

| Optical probe polarization | Second harmonic electric field | |
|---|---|---|
| | Real | Imaginary |
| $\sigma^+$ | 0 | $\chi^c_{xyz} + \chi^c_{xzy}$ |
| $\sigma^-$ | 0 | $\chi^c_{xyz} + \chi^c_{xzy}$ |

3. *SHG analytic solution for TFISH and SHG induced by time-reversal symmetry breaking*

Table S7 summarizes the expected SHG circular dichroism signals $I_{CD}$(RCP)–$I_{CD}$(LCP) as a function of φ for TFISH contributions from cubic $m3m$ and the emergent contribution from time-reversal symmetry breaking. We note that only the lowest-order (second-order) terms were considered. Both TFISH and emergent SHG signals exhibit four-fold rotational symmetry in their polar patterns. However, the emergent SHG signals show an additional phase shift due to interference between the sin(4φ) and cos(4φ) components, which originate from the time-variant and time-invariant parts of the 4/*mm*′*m*′ EQ process, respectively. These theoretical expectations are consistent with our experimental observations, as shown in Figs. 4c and 4d.

**Table S7. THz-field-induced second harmonic field with RCP and LCP probe polarizations for cubic *m*3*m* and tetragonal 4/*mm*′*m*′ point groups. Here, coefficients *A* and *B* are functions of third-order tensors $\chi^i$ and $\chi^c$ from 4/*mm*′*m*′ symmetry.**

| Configuration | SHG signal | |
|---|---|---|
| | TFISHG (*m*3*m*) | Time-reversal symmetry breaking (4/*mm*′*m*′, 4*m*′*m*′) |
| $I_{CD}$(RCP) – $I_{CD}$(LCP) | $(\chi^i_{xxxx} + \chi^i_{yyzz} - \chi^i_{yzyz} + \chi^i_{yzzy})E_0E_x - (\chi^i_{xxxx} - \chi^i_{yyzz} - \chi^i_{yzyz} - \chi^i_{yzzy})E_0E_x \cos(4\varphi)$ | $A + B\cos(4\varphi + \phi)$ |

## Section S1. Characterization of elliptically polarized THz pulse

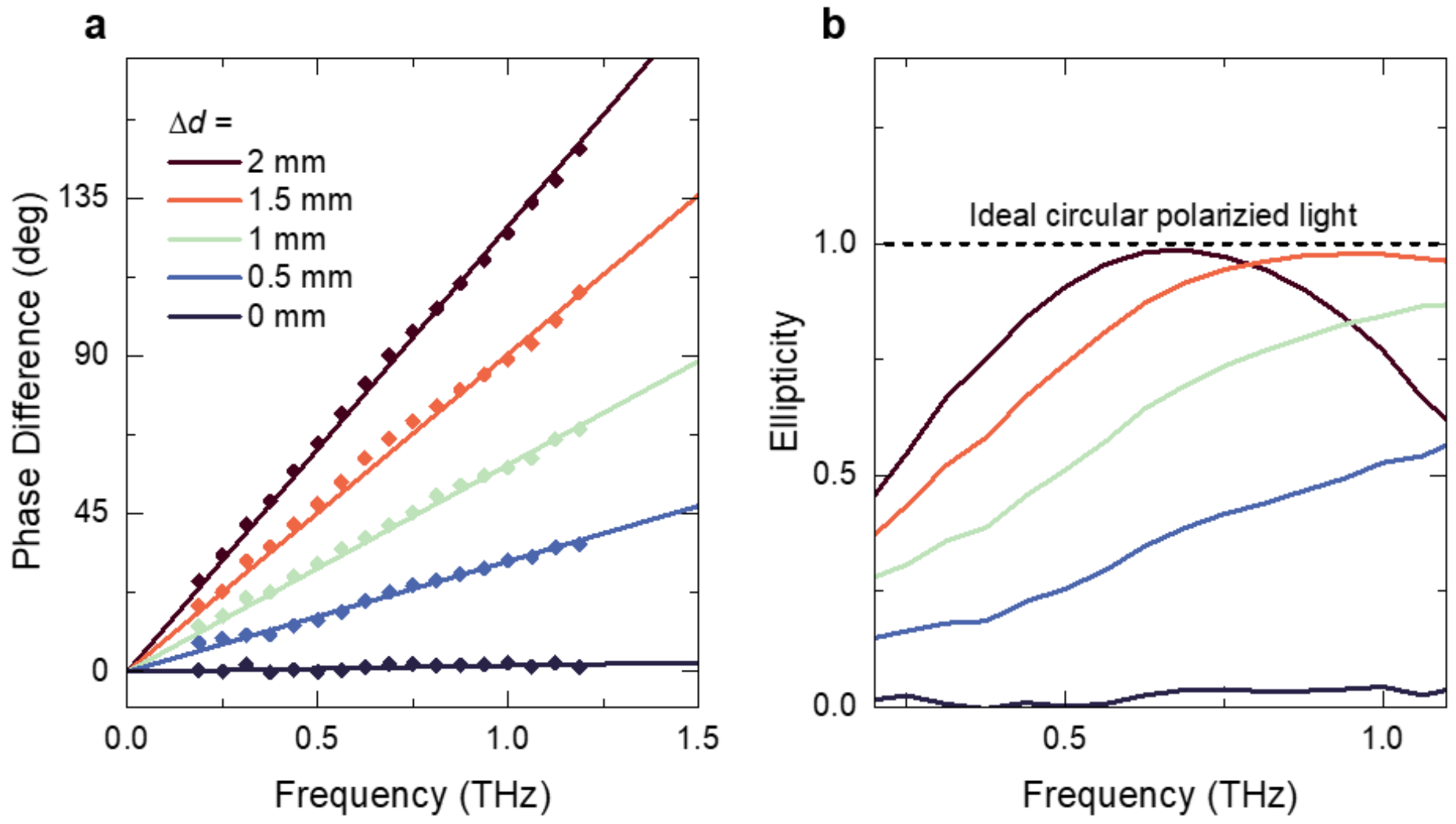

**Figure S1. Phase difference and ellipticity of the THz pulse in the frequency domain, measured at several off-center positions of the polarizer Δ*d*.**

To characterize elliptically polarized THz pulses induced by an off-centered Sagnac interferometer, we obtained FFT spectra from the time-profiles of THz pulses obtained by electro-optic sampling. A time delay $\Delta t$ between two identical pulses introduces a frequency-dependent phase shift $\omega\Delta t$, which is a direct consequence of the Fourier transform's time-shifting property. Figure S1a shows the phase difference between the *x*- and *y*-components of THz pulses in the frequency domain obtained at several $\Delta d$ values. The phase difference shows a clear linear dependence on frequency, with its slope systematically varying as a function of $\Delta d$. We calculated the ellipticity of THz pulses, which can be obtained by Stokes parameters[7], as shown in Fig. S1b. We note that the THz pulses are perfectly circular when the ellipticity equals 1. At $\Delta d$ = 2 mm, the THz pule is perfectly circular at 0.6 THz, and the ellipticity is gradually reduced as $\Delta d$ decreases. This time-delay-based method for controlling THz pulse ellipticity can be applied to any THz source without power loss and enables THz-helicity-dependent measurements with minimal error.

**Section S2. SHG polarimetry under varying polarization states of the THz pump and optical probe pulses**

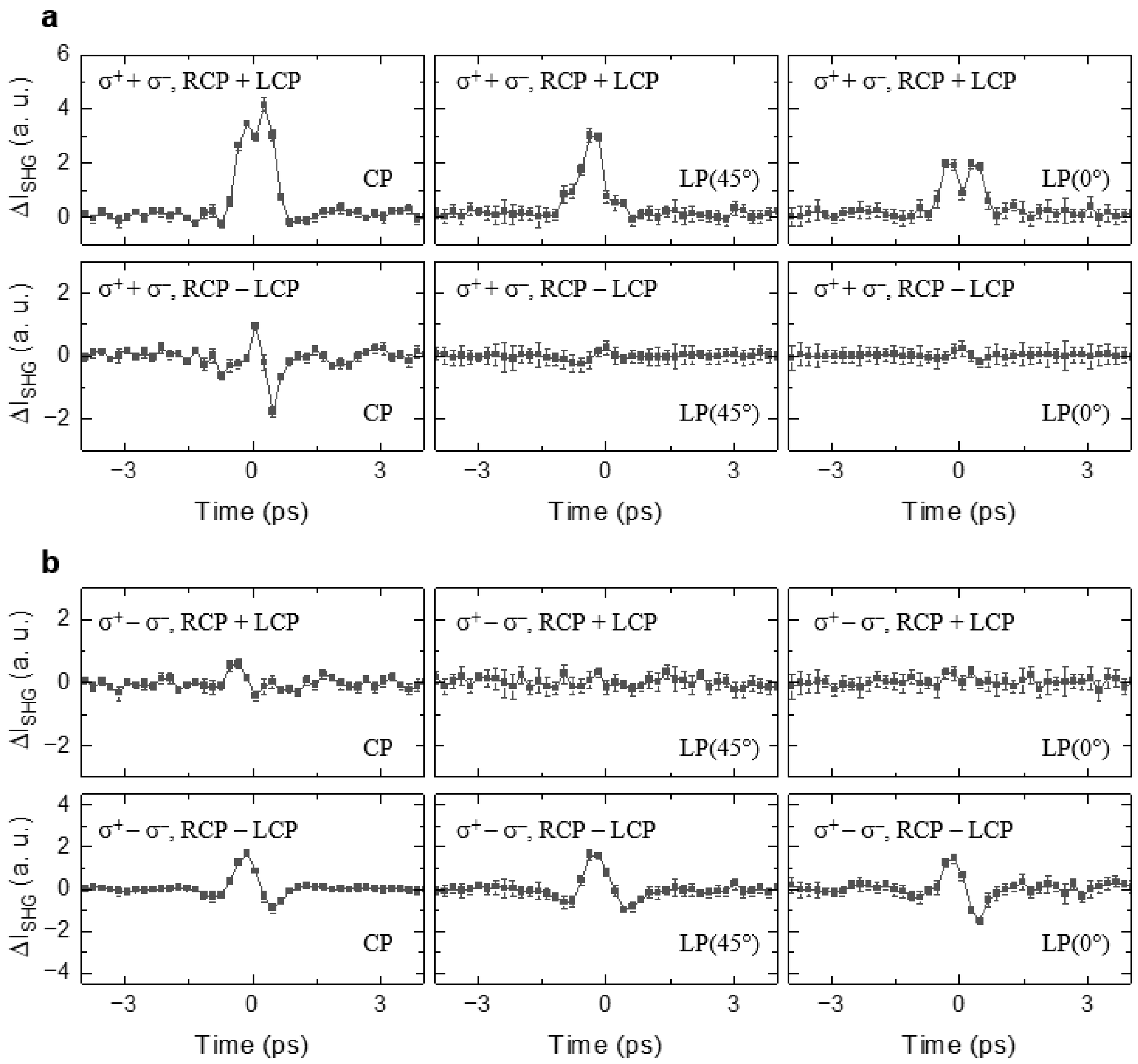

**Figure S2. Transient SHG signals $\Delta I_{SHG}$ obtained at φ = 0° with circularly polarized optical probe pulse ($\sigma^+$ and $\sigma^-$) and elliptically polarized THz pulse (RCP and LCP).**

To confirm the validity of the SHG polarimetry method, we compared all polarization combination of $\Delta I_{SHG}$ to the analytic solution, as shown in Table S2 (Supplementary Note 2). Figures S2a and S2b exhibit $\Delta I_{SHG}$ obtained at φ = 0° for four polarization combinations ($\sigma^+ + \sigma^-$, RCP ± LCP) and ($\sigma^+ - \sigma^-$, RCP ± LCP), respectively, composed of two optical-probe helicities ($\sigma^+$ and $\sigma^-$) and two THz-pump helicities (RCP and LCP). We note that CP (left panels) refers to the elliptically polarized THz pulse without using a wire-grid polarizer. In contrast, LP(45°) (middle panels) and LP(0°) (right panels) represent linearly polarized THz pulses generated from the CP by placing a wire-grid polarizer at 45° and 0°, respectively. If we considered only TFISH contributions, $\Delta I_{SHG}$ should disappear at the combinations ($\sigma^+ + \sigma^-$,

RCP – LCP) and ($\sigma^+ - \sigma^-$, RCP + LCP), as shown in the bottom of Fig. S2a and top panels of Fig. S2b, respectively. We found that LP(45°) and LP(0°) exhibit negligible SHG signals, demonstrating that TFISH contributions are dominant under linearly polarized THz excitation. On the other hand, CP exhibits weak non-zero oscillatory signals, which can be explained by the emergent SHG signals from time-reversal symmetry breaking. The other configurations ($\sigma^+ + \sigma^-$, RCP + LCP) and ($\sigma^+ - \sigma^-$, RCP – LCP) exhibit finite signals for all THz polarizations states, where the intensity contribution $|E_{x,y}|^2$ and heterodyne contribution $E_0E_x$ are dominant, respectively. These results are also consistent with our analytic solutions in Table S2, demonstrating the validity and robustness of the SHG polarimetry.

## Section S3. THz helicity-dependent magneto-optic effect

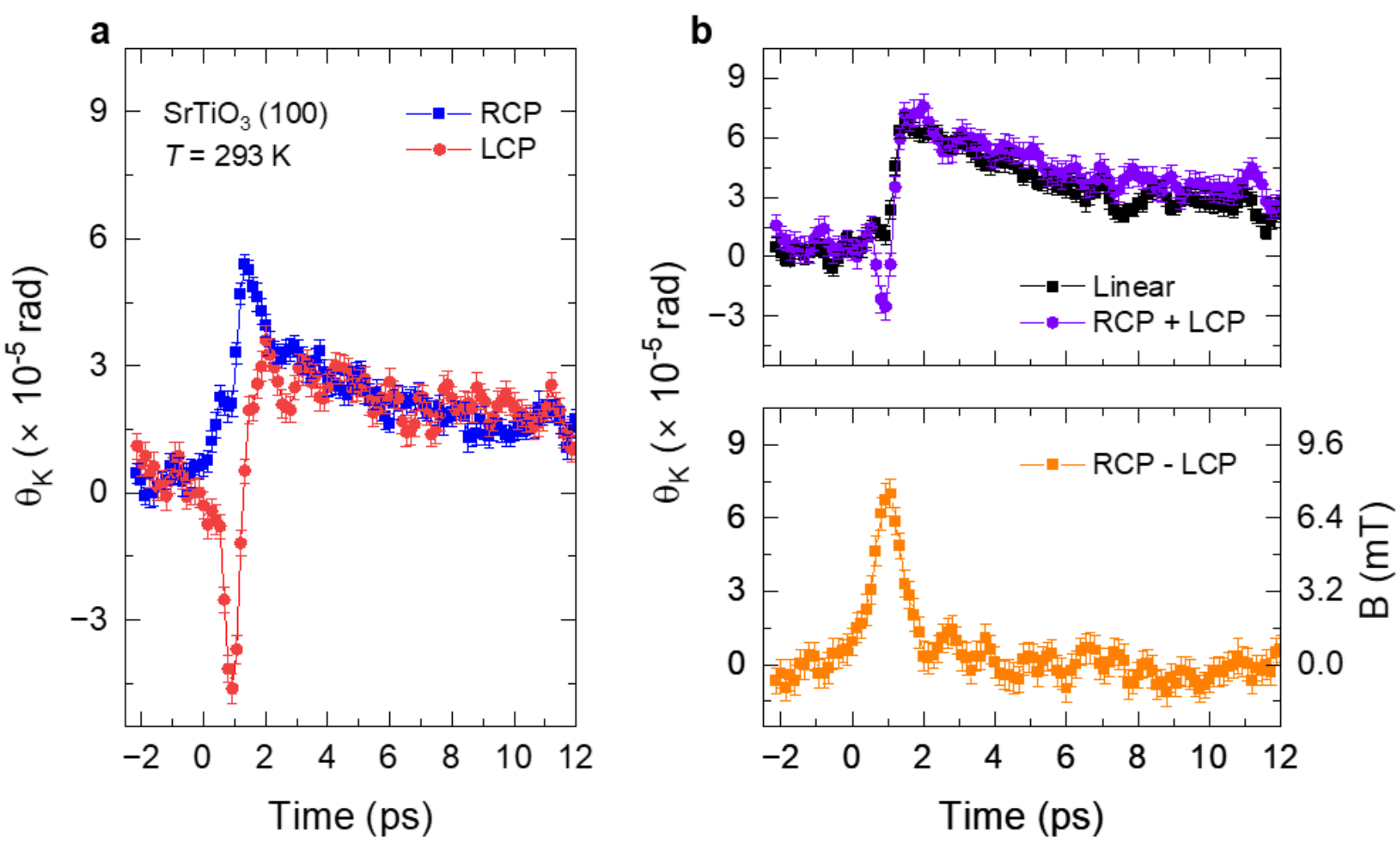

**Figure S3. Transient polarization rotation $\theta_K$ under RCP and LCP THz excitation.**

To explore THz-field-induced magnetization in STO, we conducted time-resolved magneto-optic Kerr measurements under elliptically polarized THz excitation with two helicities (RCP and LCP). The Kerr rotation angle $\theta_K$ was obtained at the normal incidence, and the probe polarization angle was set to be parallel to the [010] direction of STO. A previous study on STO has demonstrated that resonant circular THz excitation can generate axial transverse optical (TO) phonons that induce a non-oscillatory magneto-optic signal[8]. We observed a similar THz-helicity-dependent non-oscillatory magneto-optic signal obtained under RCP (blue) and LCP (red) THz excitation, as shown in Fig. S3a. In our case, the center frequency of the THz pulse (~0.5 THz) is much lower than the TO phonon frequency (~3 THz), so a resonant response is expected to be weak. Nevertheless, the observed helicity-dependent, non-oscillatory Kerr response cannot be straightforwardly assigned to an instantaneous non-linear THz Kerr effect alone. This motivates our consideration of a transient effective magnetization associated with helicity-selective excitation of circulating ionic states. Figure S3 presents the sum RCP + LCP (top, purple) and the difference RCP – LCP (bottom, orange) of $\theta_K$ obtained under RCP and LCP THz excitation. Notably, the helicity-independent RCP + LCP signal exhibits dynamics identical to those observed under linearly polarized THz excitation aligned with the probe polarization. It can be attributed to THz-field-induced birefringence resulting from the alignment of spontaneous polarization under single-cycle THz excitation. This large THz-field-induced polarization implies that we can expect an observable magnetic moment induced by dynamical multiferroicity ($\boldsymbol{m} \sim \boldsymbol{P}\times\mathrm{d}\boldsymbol{P}/\mathrm{d}t$)[9] even under off-resonant excitation. RCP – LCP shown in the bottom panel highlights the purely helicity-dependent component of the response, which exhibits a non-oscillatory signal indicating the generation of a transient magnetic moment under elliptically polarized THz excitation. We calculated the magnetic field $B$ induced by elliptically polarized THz pules from the equation $\theta_F = VBl_{decay}/2$, where $V$ is the Verdet constant, and $l_{decay}$ is the THz penetration depth, approximately 100 μm at 0.5 THz[10]. We found that the maximum $B$ value is about 8 mT, which is an order of magnitude smaller than that induced by ionic excitation[8].

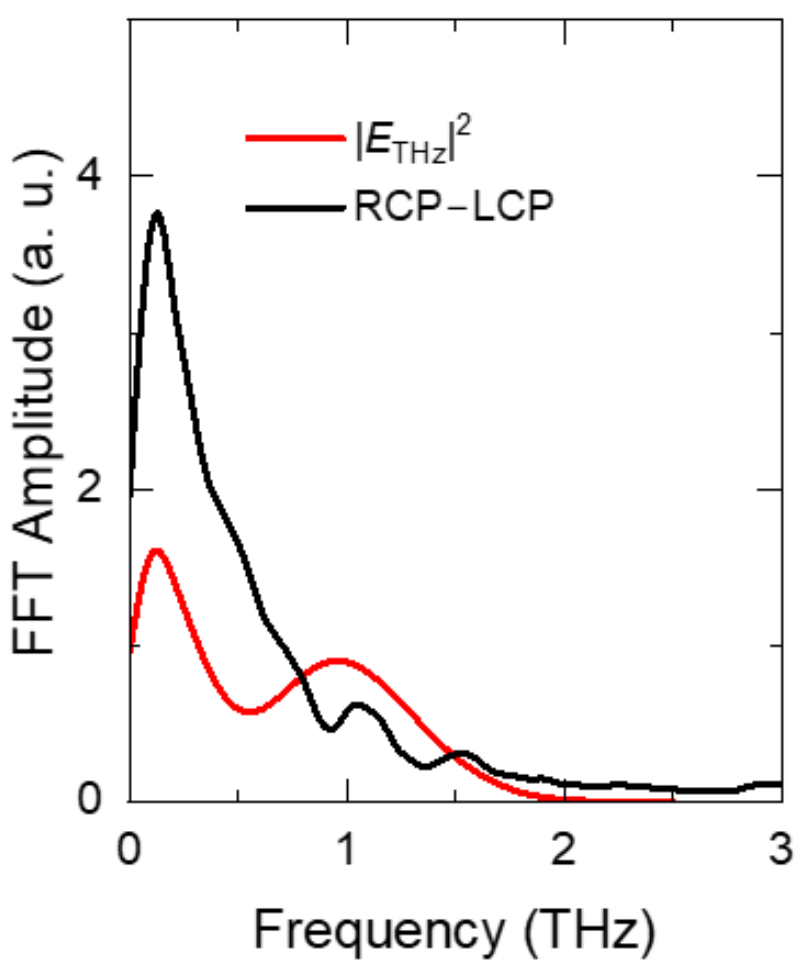

**Figure S4. FFT spectrum of the square of the THz *E*-field $|E_{THz}|^2$ (red) and THz-helicity-dependent magneto-optic signal RCP – LCP (black).**

However, we carefully interpret the THz-helicity-dependent magneto-optic signal, which can be attributed to THz electronic and ionic Kerr effect[11]. Under THz excitation, both the THz and optical probe fields can modulate the refractive index via a third-order nonlinear process, known as the THz electronic Kerr effect. THz-field-induced TO phonons can also modulate the refractive index, known as THz ionic Kerr effect; however, we can rule out their contribution because the THz frequency is lower than the TO phonon frequency, as discussed previously. Under circularly polarized THz excitation, THz electronic Kerr effect can emerge for all sample azimuth angles, resulting in non-zero transient $\theta_K$ that should follow the square of THz *E*-field[11] $|E_{THz}|^2$. Figure S4 shows the comparison of FFT spectra between $|E_{THz}|^2$ (red) and THz-helicity-dependent magneto-optic signal RCP – LCP (black). We found that their FFT spectra show a clear difference, with the non-oscillating DC component more pronounced in RCP – LCP. This suggests that the observed non-oscillatory signal in RCP – LCP arises from an effective magnetic field induced by the elliptically polarized THz pulse, rather than from the non-linear THz Kerr effect.

**Section S4. Simulation of THz-field-induced ionic magnetization**

Our observation of THz-driven ionic magnetism in STO raises an important question about the role played in this effect by the properties specific to this material. At cryogenic temperatures, STO forms a quantum paraelectric (QPE) state where the ionic displacement potential exhibits shallow minima at a finite displacement, but ferroelectric (FE) ordering is prevented by quantum effects. Below, we outline the conventional classical picture of ionic magnetism, and then develop a minimal model demonstrating that the quantum mechanisms that underlie QPE result in non-classical enhancement of ionic magnetism even at room temperature.

*Classical ionic polarization and magnetism*

In the classical approximation for THz-driven dynamics of the TO mode[9], the position of $k^{\text{th}}$ ion in a unit cell is described by a well-defined displacement $u_k$ and its contributions to the dipole moment $\boldsymbol{p_k} = Z_k^* u_k$ are also well-defined. Here, $Z_k^*$ is the Born effective charge of $k^{\text{th}}$ atom, including the electronic contribution to susceptibility in the adiabatic approximation.

To set up the framework for evaluating the dynamical characteristics relevant to THz-driven ionic dynamics, we adopt the Slater approximation for the TO mode, which assumes that the Ti cation and rigid oxygen octahedra are oppositely displaced, and neglect the Sr displacement. The mode can be visualized as an effective single ion displacement using the effective mode charge $Z^* = \sum_k \frac{Z_k^*}{\sqrt{M_k}}\sqrt{M} \approx 1.8$, where the summation over the ions excludes Sr. $M = \frac{1}{4}\sum_k M_k = 24$ amu is the effective mode mass, and the Born effective ion charges estimated from the ab initio calculations[12] are $Z_{Ti}^* \approx 7.5$, $Z_{O\|}^* \approx -5.7$, $Z_{O\perp}^* \approx -2.1$. The corresponding effective mode coordinate is equal to the relative Ti-O displacement $u = u_{Ti} - u_O$ such that the mode kinetic energy is $T = \frac{M\dot{u}^2}{2}$.

In the classical model of ionic dynamics, dipolar coupling to the THz field $\boldsymbol{E(t)}$ results in polarization dynamics described in the Lorentz model by

$$\ddot{\boldsymbol{p}} = \frac{(eZ^*)^2}{M}\boldsymbol{E}(t) - \omega_{TO}^2\boldsymbol{p} - 2\Gamma\omega_{TO}\dot{\boldsymbol{p}}, \qquad \text{(S1)}$$

where $\boldsymbol{p} = eZ^*\boldsymbol{u}$ is the dipole moment per unit cell, $\omega_{TO} = 1.8 \times 10^{13}$ s$^{-1}$ is the angular frequency of TO mode at room temperature, and $\Gamma$ is the mode damping parameter. Circularly polarized light results in polarization rotation, producing a classical magnetic moment

$$\boldsymbol{\mu_c} = \frac{\boldsymbol{p} \times \dot{\boldsymbol{p}}}{2eZ^*}. \tag{S2}$$

For the precession of $\boldsymbol{p}$ driven by circularly polarized CW field $E_x + iE_y = E_0 e^{i\omega t}$,

$$\boldsymbol{\mu_c} = \frac{\omega |\boldsymbol{p}|^2}{2eZ^*} \hat{\boldsymbol{z}}, \tag{S3a}$$

$$|\boldsymbol{p}| = \frac{\varepsilon_0 \chi_0 E_0}{\sqrt{\left(1 - \frac{\omega^2}{\omega_{TO}^2}\right)^2 + i\Gamma_c^2 \frac{\omega^2}{\omega_{TO}^2}}} \tag{S3b}$$

where $|\boldsymbol{p}|$ is the magnitude of the polarization. In the harmonic approximation Eq. S1, the magnetization is quadratic in field (linear in pump intensity).

*Breakdown of the classical ionic dynamics in STO*

The classical picture of THz-driven polarization and magnetization dynamics assumes that the displacement potential forms a narrow minimum at the centrosymmetric position $\boldsymbol{u} = 0$, such that the ionic position and its velocity that determine $\boldsymbol{p}$, $\dot{\boldsymbol{p}}$ are well-defined. However, this approximation does not hold for STO even at room temperature, as shown by the following analysis.

The shallow displacement energy landscape of incipient FE in STO results in a large ionic position uncertainty $\Delta x_0$, which is also reflected by the softening of the TO phonon mode. In the harmonic approximation, the position uncertainty due to zero-point fluctuations at room temperature is $\Delta x_0 = \sqrt{\frac{\hbar}{2M\omega_{TO}}} = \frac{l}{\sqrt{2}} \approx 8.4$ pm, where $l = \sqrt{\frac{\hbar}{M\omega_{TO}}} = 12$ pm is the oscillator length. These values are an order of magnitude larger than in typical dielectrics, reflecting the shallow displacement energy landscape. The actual value of position uncertainty is larger due to thermal fluctuations and the anharmonicity of the shallow potential of incipient FE. The former can be accounted for in the classical approximation using e.g. the Langevin approach. For the latter, we note that strain significantly enhances the dielectric constant of STO, and can even result in FE ordering at room temperature[13], indicating that the finite-displacement potential minima that exist in the QPE state may persist even at room temperature. This possibility is supported by studies of a sister compound $BaTiO_3$, which show that eight displacement energy minima located in the <111> family of directions at Ti displacement $d_{BTO}$ $\approx$ 12-17 pm from the centrosymmetric position persist in the cubic paraelectric phase[14]. For STO, the same scenario of shallow Ti displacement minima in the <111> directions at displacement of about 8-10 pm is supported by diffuse X-ray scattering[15]. This shallow-displacement-minima picture is not essential for our analysis of quantum effects in STO, as demonstrated by the mapping to the harmonic approximation. Nevertheless, it may prove important for precise quantitative modeling of ionic dynamics.

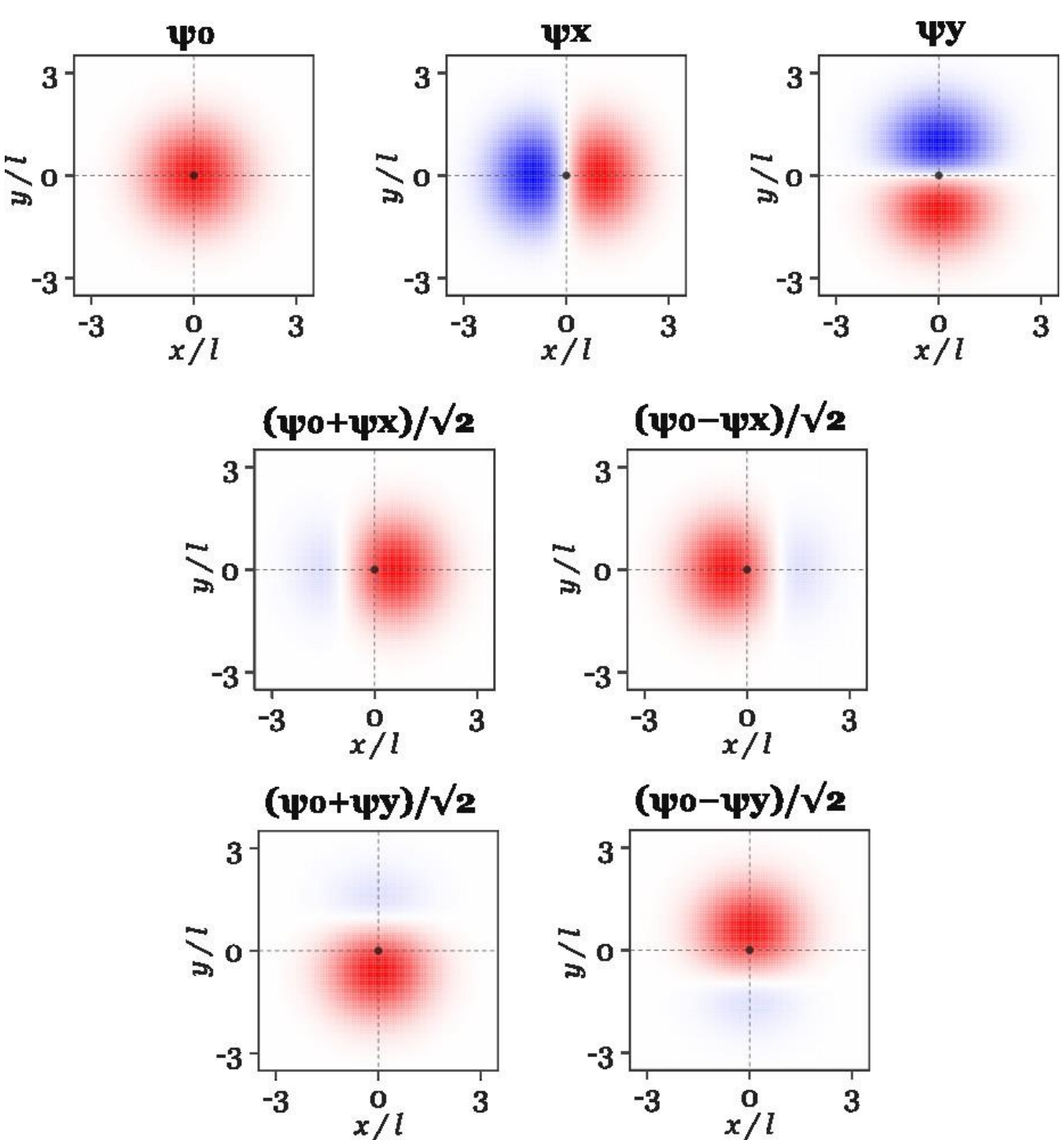

**Figure S5. Pseudo-color maps of representative wavefunctions in the harmonic approximation illustrating how displacements are described by superpositions of the basis states $\psi_0$, $\psi_x$, and $\psi_y$. Red: positive amplitude, blue: negative amplitude.**

To evaluate the significance of the position uncertainty, we compare it to field-induced displacement. In the quasi-static limit, the displacement $u$ =10 pm is reached at field $E_0 = \frac{uM\omega_{TO}^2}{eZ^*} \approx 4.6$ MV/cm. Thus, displacement becomes comparable to the position uncertainty only at very large fields, invalidating the classical approximation for ionic dynamics at fields in the 100 kV/cm to 1 MV/cm range used in THz pump experiments such as those presented in this work. Below, we examine the role of quantum effects in ionic dynamics, starting from two limiting approximations for the ionic potential: *i*) the harmonic approximation and *ii*) the tight-binding approximation assuming an incipient FE ionic potential, and we show that they are equivalent if constrained to the three lowest-energy ionic states.

*Quantum ionic model in harmonic approximation*

In the isotropic harmonic approximation, the Hamiltonian describing ionic dynamics in the *xy* plane dominated by the TO mode is

$$H = -\frac{\hbar^2\nabla^2}{2M} + \frac{M\omega_{TO}^2\boldsymbol{r}^2}{2}, \tag{S4}$$

which acts on the wavefunction $\psi(\boldsymbol{r}, t)$ describing the distribution of probability amplitude for the mode displacement $\boldsymbol{u} = \boldsymbol{r}$. Here, $M$ is the effective mass of TO mode.

We limit analysis to the three lowest-energy states, which captures the essential distinction between the classical and quantum descriptions. The population of the next excited state, while non-negligible, is substantially lower (about 3 times lower than the ground state for STO at room temperature). The three lowest-energy states are $\psi_0 \equiv \psi_{00}$, $\psi_x \equiv \psi_{10}$, and $\psi_y \equiv \psi_{01}$, where double subscripts denote the harmonic levels for the $x$ and $y$ directions, respectively. The ionic wavefunction $\psi_0$ of the ground state with energy $E_0 = \frac{1}{2}\hbar\omega_{TO}$ is isotropic and does not carry net polarization or magnetic moment. The lowest-energy excited state with energy $E_1 = \frac{3}{2}\hbar\omega_{TO}$ is two-fold degenerate. The states $\psi_{L,R} = \frac{1}{\sqrt{2}}(\psi_x \pm i\psi_y)$ characterized by angular momentum $L_{L,R} = \pm\hbar$ break time-reversal symmetry and carry circulating ionic current, and thus ionic magnetic moment

$$\mu_{L,R} = \frac{eZ^*}{2M}\langle\psi_{L,R}|\hat{L}|\psi_{L,R}\rangle = \pm m_0, \tag{S5}$$

where $\hat{L} = -i\hbar(x\partial_y - y\partial_x)$ is the z-component of angular momentum, $m_0 = \frac{eZ^*\hbar}{2M} \approx 4 \times 10^{-5}\mu_B$, and $\mu_B$ is the Bohr magneton. The dipole operator is $\hat{\boldsymbol{p}} = eZ^*\hat{\boldsymbol{r}}$.

These states are excited from the ground state by LCP and RCP light, respectively, as described by the matrix elements of the complex dipole operator $\hat{p}_\pm = \hat{p}_x \pm \hat{p}_y = \frac{eZ^*}{\sqrt{2}}(\hat{x} \pm i\hat{y})$,

$$\langle\psi_L|\hat{p}_+|\psi_0\rangle = -\langle\psi_R|\hat{p}_-|\psi_0\rangle = p_0;\ \langle\psi_R|\hat{p}_+|\psi_0\rangle = \langle\psi_L|\hat{p}_-|\psi_0\rangle = 0, \tag{S6}$$

where $p_0 = eZ^*l/\sqrt{2}$. Since $\langle\psi_i|\hat{\boldsymbol{p}}_{x,y}|\psi_i\rangle = 0$ for all the basis wavefunctions, dipole moment is carried only by the superpositions of $\psi_x$or $\psi_y$ and $\psi_0$ (see Fig. S5),

$$\langle\alpha\psi_0 + \beta_x\psi_x|\hat{p}_x|\alpha\psi_0 + \beta_x\psi_x\rangle = 2p_0 Re(\alpha^*\beta_x), \tag{S7a}$$

$$\langle\alpha\psi_0 + \beta_y\psi_y|\hat{p}_y|\alpha\psi_0 + \beta_y\psi_y\rangle = 2p_0 Re(\alpha^*\beta_y)p_0., \tag{S7b}$$

By limiting the basis to three states, the model limits the maximum dipole moment per cell to $p_0$. The nonlinearity onsets at fields comparable to $E_0$, providing a reasonable approximation for the nonlinearity of susceptibility observed at fields approaching 1 MV/cm at cryogenic temperatures[16].

The above analysis elucidates the distinction between the quantum and the classical approximations for time-reversal broken states. Consider a pure state $\psi = \alpha\psi_0 + e^{-i\omega_{TO}t}\beta\psi_L$ with real $\alpha$, $\beta$ that describes counterclockwise rotation of dipole moment excited by the LCP pump. Impure states appropriate for room-temperature analysis are discussed below. The dipole moment given by Eq. S5 is $\langle\boldsymbol{p}\rangle = \sqrt{2}p_0\alpha\beta(\cos(\omega_{\mathrm{TO}}t), \sin(\omega_{\mathrm{TO}}t))$. According to Eq.

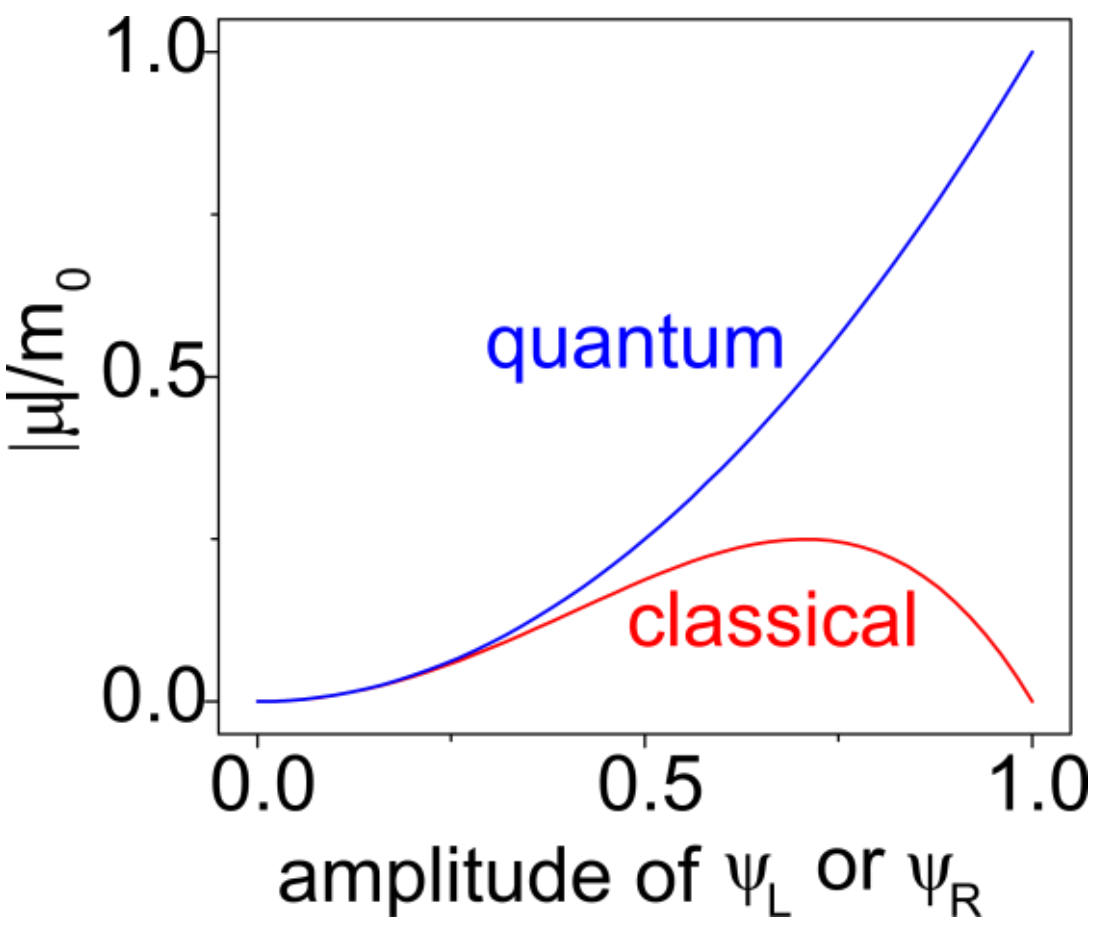

**Figure S6. Classical and quantum magnetic moments vs the amplitude β of $\psi_L$ or $\psi_R$ in a pure state $\psi = \alpha\psi_0 + \beta\psi_{L,R}$.**

S3 the classical magnetic moment is $\mu_c = \frac{\alpha^2\beta^2 p_0^2 \omega_{TO}}{eZ^*} = \alpha^2\beta^2 m_0$. According to Eq. S5, the quantum value $\mu_q = m_0\beta^2$ is always larger due to the non-classical contribution to ionic current, as illustrated in Fig. S6. The maximum classical magnetic moment described by these equations is $m_0/4$, four times smaller than the maximum quantum value $m_0$. Moreover, if a circularly-polarized pump produces an ionic state with a small amplitude α of $\psi_0$ and a large amplitude of β of $\psi_L$ or $\psi_R$ , this state does not carry polarization or classical magnetization (Eq. S2) but carries a large quantum magnetization (Eq. S5, Fig. S6).

*Quantum ionic model in the four-well tight-binding approximation*

We showed above that the harmonic approximation can describe the non-classical enhancement of magnetization in ionic dynamics analyzed in this Section. However, based on experimental observations described above, the ionic displacement landscape is more accurately approximated by eight shallow potential minima displaced in the <111> directions from the centrosymmetric position, which for displacements in the *xy* plane project onto four minima in the <11> directions. We now show that in the basis of the three lowest-energy wavefunctions, the four-well model is equivalent to the harmonic approximation.

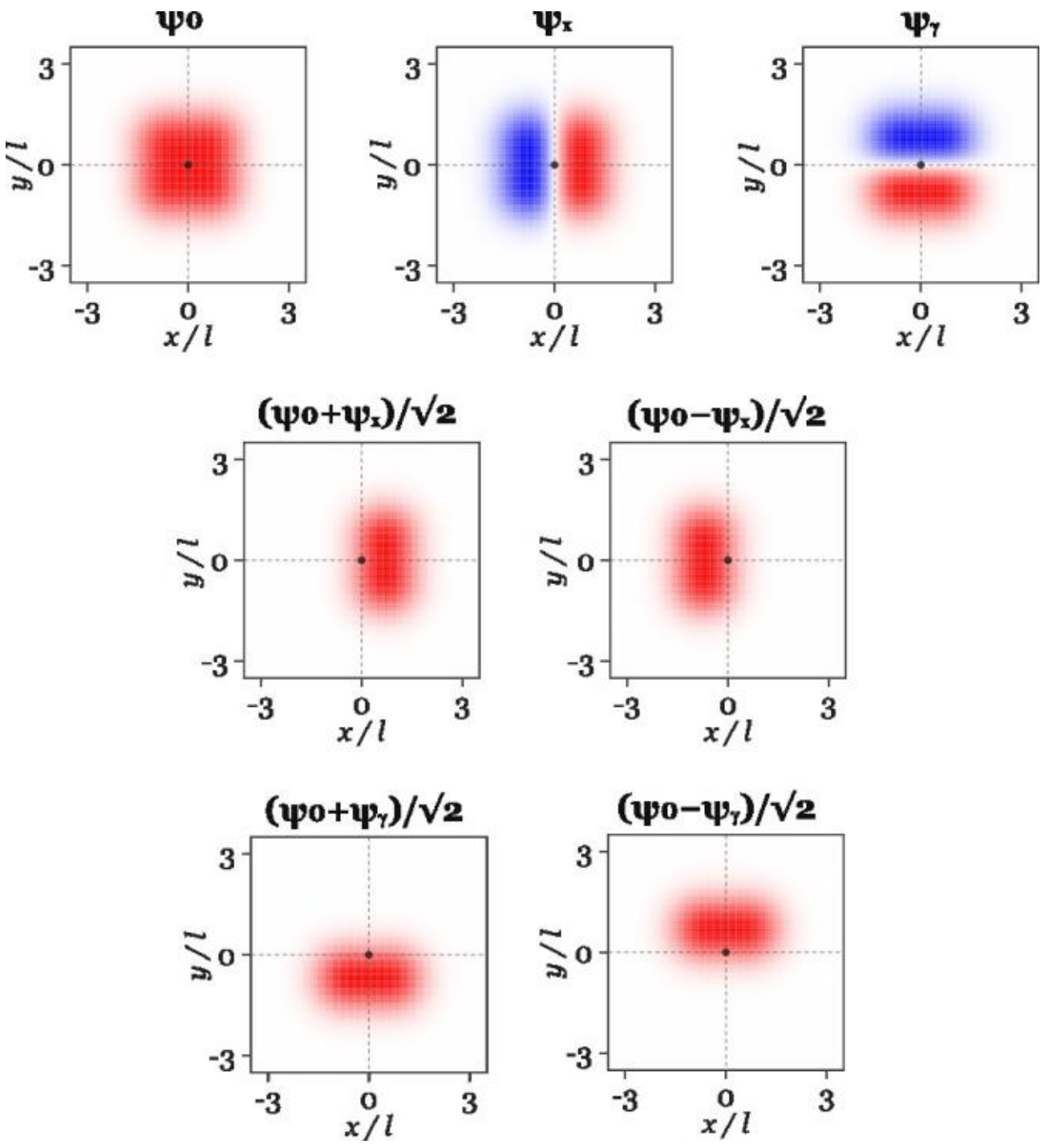

**Figure S7. Pseudo-color maps of representative wavefunctions in the four-well approximation illustrating how displacements are described by superpositions of the basis states $\psi_0$, $\psi_x$, $\psi_y$, approximated as superpositions of Gaussians with half-width $\sigma$ = 2/3. Red: positive amplitude, blue: negative amplitude.**

In the tight-binding approximation, the lowest-energy states of the four-well potential are described by superpositions of basis wavefunctions $\phi_0$, $\phi_1$, $\phi_2$, and $\phi_3$ centered at the potential minima at positions $(d, d)$, $(-d, d)$, $(-d, -d)$ and $(d, -d)$, respectively, in the *xy* plane. We evaluate the value of *d* below by matching the oscillator length to a harmonic approximation. The matrix elements of the Hamiltonian

$$\langle\phi_j|H|\phi_{j\pm1}\rangle = -V/2, \tag{S8}$$

with $V > 0$ describing hybridization between the neighboring minima. The ground state $\psi_0 = \frac{1}{2}\sum_{j=0}^{3}\phi_j$ of this Hamiltonian with energy $E_0 = -4V$ is symmetric under the $C_4$ rotation symmetry of the system. The two lowest-energy excited states $\psi_{L,R} = \frac{1}{2}\sum_{j=0}^{3} e^{\pm i\pi(j+\frac{1}{2})/2}\phi_j$ are degenerate with energy $E_1 = 0$, transforming under the right- and left-handed representations

$E_{1,2}$ of the $C_4$ symmetry, respectively. Alternatively, one can use the displacement basis functions

$$\psi_x = \frac{1}{2}(\phi_0 - \phi_1 - \phi_2 + \phi_3), \quad \text{(S9a)}$$

$$\psi_y = \frac{1}{2}(\phi_0 + \phi_1 - \phi_2 - \phi_3). \quad \text{(S9b)}$$

For the three lowest-energy basis states $\psi_0$, $\psi_x$, $\psi_y$, the four-well model maps directly to the harmonic oscillator approximation, as also illustrated by the close similarity between the spatial profiles of the corresponding wavefunctions (compare Fig. S7 to Fig. S5).

Mapping between the energies and the oscillator lengths of the two models gives $V = \frac{1}{4}\hbar\omega_{TO}$ and $d = \frac{l}{\sqrt{2}}$. For the latter, we for simplicity neglect the corrections due to the finite spatial extent of the quasi-localized basis wavefunctions. The dipole moment is given by the same expressions Eqs. S7 as in the harmonic model. The exact value of the magnetic moment $\mu_{L,R} = \frac{eZ^*}{2Mc}\langle\psi_{L,R}|\hat{L}|\psi_{L,R}\rangle$ depends on the spatial structure of $\psi_{L,R}$, giving corrections to the harmonic value Eq. S5 of order 1, as also illustrated by the similarity of the corresponding wavefunctions (Fig. S8).

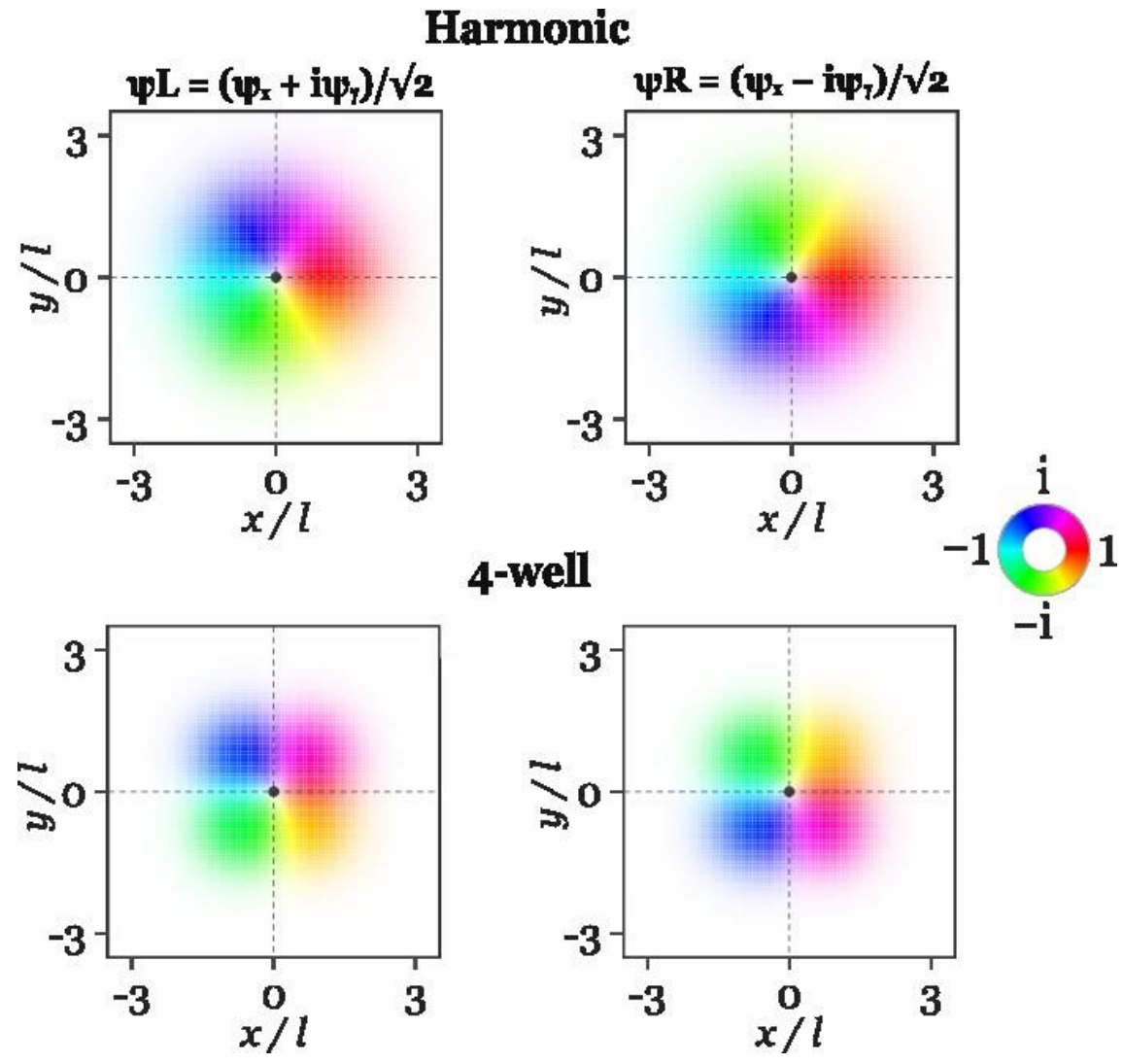

**Figure S8. Pseudo-color maps of wavefunctions $\psi_L$ (left), $\psi_R$ (right) in the harmonic (top) and four-well (bottom) approximations, with the latter approximated as a superposition of Gaussians with half-width $\sigma = l/1.5$. Color wheel shows the mapping of the wavefunction phases.**

*Quantum model of ionic dynamics.*

In the basis of states $\psi_0$, $\psi_x$, $\psi_y$ defined either in the harmonic or the four-well approximation as described above, the Hamiltonian including the dipolar coupling to the THz field is

$$H = \begin{bmatrix} -\hbar\omega_{\mathrm{TO}} & -p_0 E_x(t) & -p_0 E_y(t) \\ -p_0 E_x(t) & 0 & 0 \\ -p_0 E_y(t) & 0 & 0 \end{bmatrix}. \tag{S10}$$

The dipole moment operators given by Eqs. S7 are

$$p_x = \begin{bmatrix} 0 & p_0 & 0 \\ p_0 & 0 & 0 \\ 0 & 0 & 0 \end{bmatrix},\ p_y = \begin{bmatrix} 0 & 0 & p_0 \\ 0 & 0 & 0 \\ p_0 & 0 & 0 \end{bmatrix}, \tag{S11}$$

and the *z*-component of the (quantum) magnetic moment is

$$\mu_q = \begin{bmatrix} 0 & 0 & 0 \\ 0 & 0 & im_0 \\ 0 & -im_0 & 0 \end{bmatrix}. \tag{S12}$$

To analyze THz-driven dynamics, we approximated the THz pump pulse by a sinusoidal function convolved with a Gaussian, $E_{x,y} = E_{0\ x,y}\cos(\Omega t - \varphi_{x,y})e^{-\left(\frac{t\Omega}{\sigma}\right)^2/2}$, with the parameters $E_{0,x} = \pm 0.7 E_{0,y}$ for RCP/LCP pulses, $\Omega = 3.3$ ps$^{-1}$, $\varphi_x = -0.25$, $\varphi_y = 1.37$, and $\sigma = 2.1$ determined by fitting the temporal field profiles that were determined experimentally through electro-optic sampling with parameter uncertainties of 5-10%. The modeled pulse peaks at $t = 0$, with the amplitude decreasing by 90% at $|t| = 1.4$ ps.

To confirm the significance of non-classical contributions to THZ-driven magnetization at room temperature, we analyze quantum dynamics using the Lindblad master equation

$$\frac{d\rho}{dt} = -\frac{i}{\hbar}[H,\rho] + D[\rho], \tag{S13}$$

where $\rho = |\psi_i\rangle\langle\psi_i|$ is the density matrix in the basis $\psi_0$, $\psi_x$, $\psi_y$ and $D[\rho] = -\Gamma\omega_{\text{TO}}(\rho - \rho_0)$ is the dissipation function accounting for the relaxation toward equilibrium $\rho_0 = e^{-\beta H}/Z$ in the Landau-Khalatnikov relaxation-time approximation. Here, $\Gamma$ is the damping parameter, and the partition function is $Z = \sum_i e^{-E_i/k_B T} = 2 + e^{\hbar\omega_{\text{TO}}/k_B T} = 3.6$ . The polarization is shown by Kerr measurements to closely follow the driving THz pulse indicating an overdamped regime, which can be attributed to dephasing due to interactions among multiple dynamical modes modeled using a critical damping parameter $\Gamma = 1$.

The density matrix provides a connection between the classical and quantum approximations. The diagonal elements $\rho_{00}$, $\rho_{11}$ and $\rho_{22}$ represent the probabilities to find the system in the corresponding quantum states, while the off-diagonal elements describe coherences between these states that determine the dipole moment $\boldsymbol{p} = 2p_0\big(Re(\rho_{01}), Re(\rho_{02})\big)$ and the $z$-component of the magnetic moment $\mu_q = 2m_0 Im(\rho_{12})$. One can see that the dipole moments and the magnetization are determined by different coherences: the former is determined by the coherence between the ground state and the excited states, while the latter by the coherence between the excited states. For an arbitrary impure quantum state, these coherences are not directly related, resulting in the quantum value of magnetization that is not directly determined by the polarization. This is demonstrated above for the pure states at large driving fields. As will be shown below, the difference in the quantum-coherent origins between classical and quantum expressions results in a substantial enhancement of quantum magnetization by almost an order of magnitude for STO at room temperature.

To elucidate the dynamics described by Eq. S13, we consider some limiting cases. For the dynamics of the $x$-component of polarization

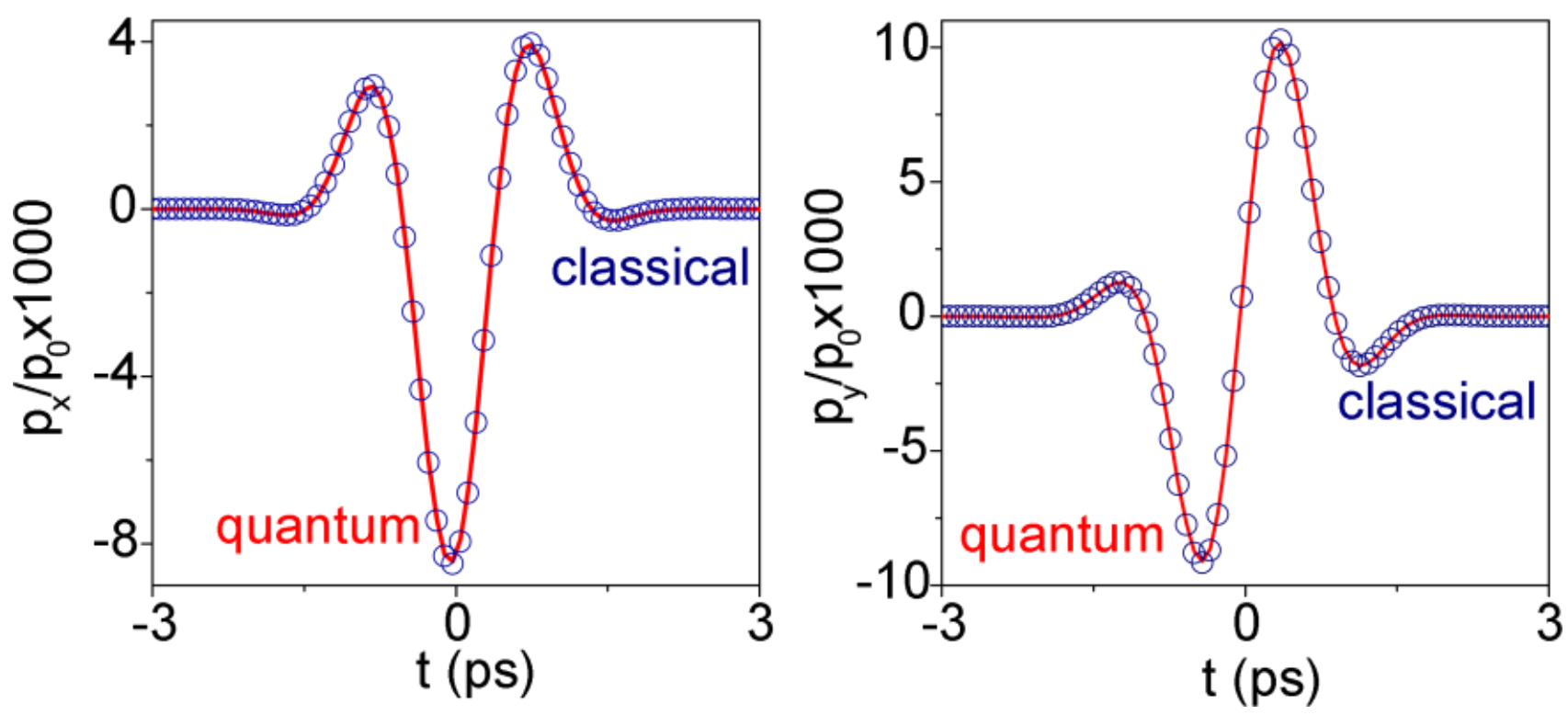

**Figure S9.** ***x*- and *y*- components of polarization vs time driven by an RCP THz pulse centered at *t* = 0 with peak amplitude 300 kV/cm, calculated using the classical approximation Eq. S2 (blue) and Lindblad Eq. S13 (red), with matched parameters as described in the text.**

$$\frac{d\rho_{01}}{dt} = i\omega_{\mathrm{TO}} - \frac{ip_0E_x}{\hbar}(\rho_{00}-\rho_{11}) + \frac{ip_0E_x}{\hbar}\rho_{21} - \Gamma\omega_{\mathrm{TO}}\rho_{01}. \tag{S14}$$

For simplicity, we assume $E_y = 0$. Separating the real from the imaginary components and differentiating, we obtain for $p_x = 2p_0Re(\rho_{01})$

$$\ddot{p}_x = -\omega_{\mathrm{TO}}^2 p_x + \frac{2p_0^2E_x\omega_{\mathrm{TO}}}{\hbar}(\rho_{00}-\rho_{11}) - \Gamma\omega_{TO}[\dot{p}_x - 2\omega_{\mathrm{TO}}Im(\rho_{01})], \tag{S15}$$

This expression reduces to the classical dynamics Eq. S1 by scaling the driving field in the classical model by the factor $w = \rho_{00}-\rho_{11} \approx 0.16$, approximating $\omega_{\mathrm{TO}}\mathrm{Im}(\rho_{01}) \approx Re(\dot{\rho}_{01})$, which is valid under weak driving, and using the identity $\frac{2p_0^2}{\hbar} = \frac{(eZ^*)^2}{M}$. As Fig. S9 illustrates, the polarization dynamics in the classical approximation track very closely the quantum result.

For the dynamics of the component $\rho_{12}$ describing the magnetization, we obtain

$$\frac{d\rho_{12}}{dt} = -\frac{ip_0}{\hbar}(E_x\rho_{02} - E_y\rho_{10}) - \Gamma\omega_{TO}\rho_{12}, \tag{S16}$$

whose imaginary part gives

$$\frac{d\mu_z}{m_0 dt} = -\frac{1}{\hbar}\left(E_x p_y - E_y p_x\right) - \frac{\Gamma\omega_{TO}}{m_0}\mu_z. \tag{S17}$$

To compare to classical magnetization dynamics, we differentiate Eq. S2 with respect to time,

$$\frac{d\mu_c}{dt} = \frac{\boldsymbol{p}\times\ddot{\boldsymbol{p}}}{2eZ^*}. \tag{S18}$$

Replacing $\ddot{\boldsymbol{p}}$ with the classical-limit Eq. S1, we obtain

$$\frac{d\mu_c}{dt} = -\frac{m_0 w}{\hbar}\left(E_x p_y - E_y p_x\right) - 2\Gamma\omega_{TO}\mu_c. \tag{S19}$$

This expression is similar to the quantum result Eq. S17, with two distinctions. First, the classical-approximation relaxation is twice as fast, because both $\boldsymbol{p}$ and $\dot{\boldsymbol{p}}$, whose product determines $\mu_c$, relax at a rate $\Gamma\omega_{TO}$. In contrast, $\mu_c$ is linear in the coherence amplitude $\rho_{12}$, which decays at a rate $\Gamma\omega_{TO}$.

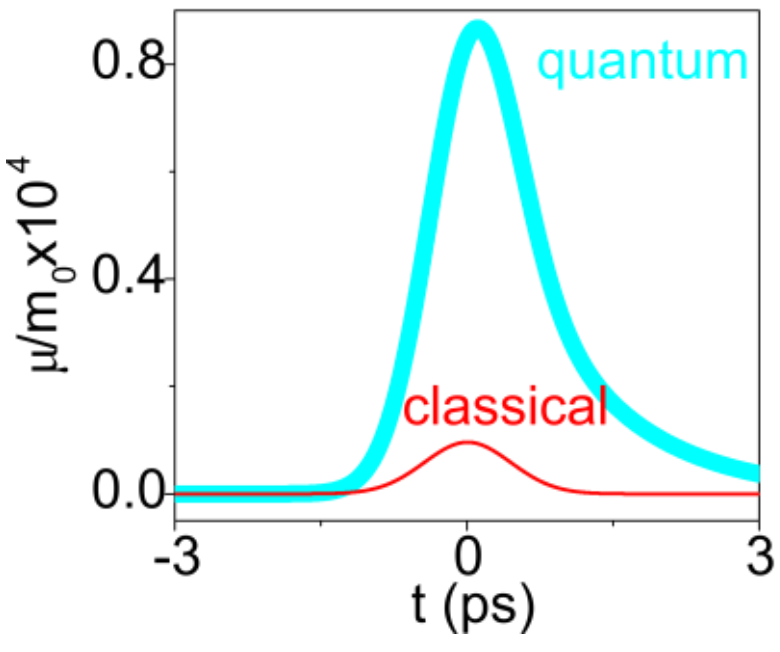

**Figure S10. Quantum vs classical ionic magnetic moment generated under the same conditions as in Fig. S9.**

Second, the effects of electric field on the classical magnetization are scaled by a factor $w \approx 0.16$, about 8 times smaller than the quantum result. This dramatic rescaling is a consequence of the anomalously low frequency of the soft TO mode in STO, since

$$w = \rho_{00} - \rho_{11} = \frac{1}{Z}\left(e^{\frac{\hbar\omega_{TO}}{k_B T}} - 1\right) \approx \frac{\hbar\omega_{TO}}{Zk_B T}. \qquad \text{(S20)}$$

For a typical oxide dielectric with $f_{\mathrm{TO}} \approx 10 - 15$ THz, $w \approx 0.6 - 0.8$; the distinction between quantum and classical effects is less significant.

We note that all the coherence factors in the above calculations are significantly smaller than 1, and therefore the nonlinear effects discussed above for the pure states and illustrated in Fig. S6 are negligible. We have separately verified this by performing calculations at different field strengths. Resonant driving with fields of the order 1 MV/cm at cryogenic temperatures, where $w \approx 1$, is expected to result in additional nonlinear quantum enhancement illustrated in Fig. S6.

Collectively, our calculations demonstrate that even non-resonant elliptically-polarized THz pulses produce non-classical dynamical ionic states in STO which cannot be described by the evolution of the classical polarization, providing a tentative explanation for the large time-reversal symmetry breaking effects observed in our SHG and magnetic Kerr effect measurements.